%% file: main.tex
\pdfoutput=1

\documentclass[sigplan,10pt,natbib=false,dvipsnames,review=false,authorsperrow=4]{acmart}

\usepackage{xspace}
\usepackage{listings}
\usepackage{subcaption}
\usepackage{amsmath}
\usepackage{fancyvrb,verbatim}
\usepackage{booktabs}
\usepackage{times}

\usepackage{ifthen}

\usepackage[xcolor]{fixedchangebar}

\newboolean{draft}
\setboolean{draft}{false}

\newboolean{extendedtr}
\setboolean{extendedtr}{true}

\newcommand{\extendedtr}[2]{%
\ifextendedtr
\begin{changebar}%
#1
\end{changebar}%
\else
#2
\fi
}

\newcommand{\fullversion}{our extended technical report\xspace} 

\usepackage[utf8]{inputenc}
\usepackage[utf8]{luainputenc}

\usepackage{nicefrac}

\usepackage{textcomp}

\usepackage{url}

\usepackage[
  backend=bibtex,
  style=numeric-comp,
  minalphanames=3,
  isbn=false,
  sortcites=true,
  sorting=anyt,
  abbreviate=true,
  url=false,
  doi=false,
  maxnames=6,
  minbibnames=7,
  maxbibnames=7]{biblatex}

\usepackage[inline]{enumitem}
\usepackage[compact, small]{titlesec}

\usepackage{etoolbox}
\makeatletter
\patchcmd{\ttlh@hang}{\parindent\z@}{\parindent\z@\leavevmode}{}{}
\patchcmd{\ttlh@hang}{\noindent}{}{}{}
\makeatother

\usepackage{balance}

\usepackage{datetime}

\usepackage{listings}

\usepackage{tikz}
\usetikzlibrary{calc}
\usetikzlibrary{trees}
\usetikzlibrary{positioning}
\usetikzlibrary{arrows}
\usetikzlibrary{chains, matrix}
\usetikzlibrary{shapes, shapes.geometric, shapes.symbols}
\usetikzlibrary{decorations.pathreplacing, decorations.pathmorphing, decorations.text}
\usetikzlibrary{backgrounds}

\usepackage{amsfonts}
\usepackage{amsmath}
\usepackage{amssymb}
\usepackage{pifont}

\usepackage{xfrac}

\usepackage{hyperref}
\hypersetup{
    colorlinks,
    linkcolor={red!50!black},
    citecolor={blue!50!black},
    urlcolor={blue!80!black}
}

\usepackage{rotating}

\usepackage{outlines}

\usepackage{xspace}

\usepackage{framed}

\usepackage{setspace}

\usepackage{comment}

\usepackage[labelfont=bf]{caption}

\usepackage{subcaption}

\lstdefinelanguage{conclave}{
  morekeywords={newTable,as,concat,aggregate,project,join,divide,multiply,writeToCSV,Column,Party,import,set,secretShare,revealTo,lookUpByIndex,concatCols,shuffle,enumerate,def,Relation,return,select,for,in,append},
  morekeywords={INT,SUM,COUNT},
  morekeywords=[2]{at,to,trust},
  sensitive=true,
  morecomment=[l]{\#},
  morecomment=[s]{\"\"\"}{\"\"\"},
  keywordstyle=\color{blue}\textbf,
  keywordstyle=[2]\color{red}\textbf,
  identifierstyle=\color{black},
  stringstyle=\color{gray},
  morestring=[b]",
}

\newcommand\todo[1]{\ifdraft{\color{red}{\bf ToDo: #1}}\fi}
\newcommand{\mv}[1]{\todo{MV: #1}}

\newcommand{\nv}[1]{\todo{NIK: #1}}

\newcommand\code[1]{\lstinline{#1}}

\newcommand\withlineno[2]{\lstinline[language=conclave]{#2}}

\NewDocumentCommand{\rot}{O{45} O{1em} m}{\makebox[#2][l]{\rotatebox{#1}{#3}}}

\DeclareMathAlphabet{\mathcal}{OMS}{cmsy}{m}{n}

\newcommand{\eg}{\textit{e.g.},\xspace}
\newcommand{\ie}{\textit{i.e.},\xspace}
\newcommand{\viz}{\textit{viz.},\xspace}

\newcommand{\one}{{\em (i)}\/}
\newcommand{\two}{{\em (ii)}\/}
\newcommand{\three}{{\em (iii)}\/}

\newcommand{\Adv}{\ensuremath{\mathcal{A}}\xspace}
\newcommand{\Sim}{\ensuremath{\mathcal{S}}\xspace}
\newcommand{\Corrupt}{\ensuremath{\mathcal{C}}\xspace}
\newcommand{\view}{\ensuremath{\text{view}^{\pi}_{\Corrupt}}}
\newcommand{\out}{\ensuremath{\text{out}^{\pi}}}
\newcommand{\Inputs}{\ensuremath{\vec{x}}\xspace}
\newcommand{\CorrInputs}{\ensuremath{{\Inputs}_{\Corrupt}}\xspace}
\newcommand{\UncorrLengths}{\ensuremath{\vec{\ell}}\xspace}
\newcommand{\Structure}{\ensuremath{\Delta}\xspace}
\newcommand{\Propagate}{\ensuremath{\mathcal{I}}\xspace}
\newcommand{\abs}[1]{\ensuremath{\vert #1 \vert}}
\newcommand{\piF}{\ensuremath{\pi^f}\xspace}
\newcommand{\piG}{\ensuremath{\pi^g}\xspace}
\newcommand{\piGF}{\ensuremath{\pi^{g | f}}\xspace}

\newcommand{\sys}{Conclave\xspace}

\newcommand\para[1]{\textbf{#1}}

\copyrightyear{2019}
\acmYear{2019}
\acmConference[EuroSys'19]{Fourteenth EuroSys Conference 2019}{March 25--28, 2019}{Dresden, Germany}
\acmBooktitle{Fourteenth EuroSys Conference 2019 (EuroSys'19), March 25--28, 2019, Dresden, Germany}
\acmPrice{15.00}
\acmDOI{10.1145/3302424.3303982}
\acmISBN{978-1-4503-6281-8/19/03}

\ifextendedtr
\setcopyright{none}
\renewcommand\footnotetextcopyrightpermission[1]{}
\else
\setcopyright{acmlicensed}
\fi

\title{\sys: secure multi-party computation on big data}
\begin{document}
\ifextendedtr
\subtitle{Extended Technical Report}
\titlenote{
\extendedtr{%
In this extended technical report version of the \sys paper~\cite{conclave},
we include additional material on \sys's security guarantees, including full
proofs in the Appendix.
A gray bar in the margin highlights the differences.
}{}%
}
\else
\fi



\author{Nikolaj Volgushev}
\affiliation{Alexandra Institute}
\authornote{Part of the work completed at Boston University.}

\author{Malte Schwarzkopf}
\affiliation{MIT CSAIL}

\author{Ben Getchell}
\affiliation{Boston University}
\author{Mayank Varia}
\affiliation{Boston University}
\author{Andrei Lapets}
\affiliation{Boston University}
\author{Azer Bestavros}
\affiliation{Boston University}

\renewcommand{\shortauthors}{N.\ Volgushev, M.\ Schwarzkopf, B.\ Getchell et al.}





\input{abstract}

\extendedtr{
\settopmatter{printfolios=true}
}{
\settopmatter{printfolios=false}
}
\maketitle


\input{intro}
\input{background}
\input{paradigm}
\input{system}

\input{impl}
\input{eval}
\input{related}
\input{concl}
\input{acks}
\input{appendix}


{
\balance


\printbibliography
}

\end{document}

%% file: abstract.tex
\begin{abstract}
%
Secure Multi-Party Computation (MPC) allows mutually distrusting parties
to run joint computations without revealing private data.
%
%
%
Current MPC algorithms scale poorly with data size, which makes
MPC on ``big data'' prohibitively slow and inhibits its practical use.

%
%
Many relational analytics queries can maintain MPC's end-to-end
security guarantee without using cryptographic MPC techniques for all
operations.
\sys is a query compiler that accelerates such queries by transforming
them into a combination of data-parallel, local cleartext processing
and small MPC steps.
When parties trust others with specific subsets of the data, \sys applies
new hybrid MPC-cleartext protocols to run additional steps outside of MPC
and improve scalability further.
%
%

%
Our \sys prototype generates code for cleartext processing in Python and
Spark, and for secure MPC using the Sharemind and Obliv-C frameworks.
%
%
\sys scales to data sets between three and six orders of magnitude larger
than state-of-the-art MPC frameworks support on their own.
Thanks to its hybrid protocols and additional optimizations, \sys also substantially outperforms
SMCQL, the most similar existing system.
\end{abstract}

%% file: intro.tex
\section{Introduction}
\label{s:intro}

Many businesses run analytics over ``big data'' to draw insights or to satisfy
regulatory requirements.
Such computations cannot currently combine proprietary, private
data sets from multiple sources, whose sharing is restricted by law and by
justifiable privacy concerns.
However, running joint analytics over large private data sets is valuable:
for example, drug companies, medical researchers, and hospitals can benefit
from jointly measuring the incidence of illnesses without revealing private
patient data~\cite{opaque, djoin, smcql}; banks and financial regulators
can assess systemic risk without revealing their private
portfolios~\cite{dstress, systemic-risk-measures}; and antitrust regulators
can measure monopolies using companies' revenue data.
Secure Multi-Party Computation (MPC) is a class of cryptographic techniques
that allow for secure computation over sensitive data sets.
In MPC, a set of participants compute jointly over the data they hold
individually, while federating trust among the computing parties.
As long as some parties\,---\,typically a majority, or any one
party\,---\,honestly follow the protocol, no party learns anything beyond
the final output of the computation.
In particular, MPC protects input and intermediate data, as well as
meta-data about them, such as value frequency distributions.
Unfortunately, implementing a performant secure MPC currently requires
domain-specific expertise that makes it impractical for most data analysts.
Moreover, existing algorithmic techniques and software frameworks for secure
MPC scale poorly with data size, even for small numbers of parties
(\S\ref{s:bg}).
These limitations have led researchers to build oblivious query processors
that generate and execute secure query plans~\cite{opaque, smcql}.
These query processors improve the accessibility of MPC, as they allow data
analysts to write convenient relational queries;
they also improve MPC
performance by securely doing part of the computation outside of MPC.
For example, SMCQL~\cite{smcql} performs local preprocessing and ``slices'' the
overall MPC into several smaller MPCs to reduce input data size, while
Opaque~\cite{opaque} relies on Intel SGX hardware to compute securely
``in-the-clear''.
%
%
%


%
This paper presents \sys, an MPC-enabled query compiler that makes MPC on ``big
data'' accessible and efficient.
Data analysts write relational queries as if they had access to all parties'
data in the clear, and \sys turns the queries into a combination of efficient,
local processing steps and secure MPC steps.
As a result, \sys delivers results with near-interactive response
times\,---\,\ie a few minutes\,---\,for input data several orders of magnitude
larger than existing systems can support.
Like prior systems~\cite{smcql,djoin,graphsc}, \sys is designed to withstand
passive (``semi-honest'') adversaries.

Three key ideas help \sys scale.
First, \sys analyses the queries to apply transformations that reduce runtime
without compromising security guarantees, even though they burden individual
parties with extra local work.
Second, \sys uses optional, coarse-grained annotations on input relations to
apply new, \emph{hybrid protocols} that combine clear-text and MPC processing
to gain further speedups especially for operations that are notoriously slow
under MPC, such as joins and grouped aggregations.
Third, \sys generates code that combines scalable, insecure data-processing
systems (\eg Spark~\cite{spark}) with secure, but slow, cross-party MPC
systems (\eg Sharemind~\cite{sharemind} or Obliv-C~\cite{obliv-c}).
\sys's optimizations are largely complementary to those introduced by
SMCQL~\cite{smcql}, the most similar related system.
\sys introduces new hybrid protocols that improve the efficiency of joins
and aggregations over private key columns in a server-aided
setting \cite{DBLP:journals/iacr/KamaraMR11,DBLP:conf/ccs/KamaraMR12}, as
well as additional query transformations that speed up the MPC steps.
Moreover, \sys generates code for both garbled-circuit and secret-sharing
MPC frameworks, as well as for data-parallel local processing systems.
Secret-sharing MPC backends are better suited to the arithmetic operations
prevalent in relational queries, and thus outperform SMCQL's
garbled-circuit backend; data-parallel local processing allows \sys to
support larger inputs.
%

%
%

%

%
%
%
%

%
In summary, this paper makes four key contributions:
\begin{enumerate}[nosep]
 %
 %
 \item query analysis techniques that derive which parts of a relational
   query must be executed under MPC using only coarse-grained annotations
   (\S\ref{s:specification}, \S\ref{s:design-propagation});
 \item transformations that move expensive oblivious operations outside the
   MPC while preserving security guarantees (\S\ref{s:design-pushupdown},
   \S\ref{s:design-min-obliv});
 \item new hybrid MPC--cleartext protocols that improve performance of
   MPC joins and aggregations using existing partial trust between parties
   (\S\ref{s:design-hybrid}); and
 \item our \sys prototype, which applies these ideas to
   generate efficient code for execution using Python,
   Spark~\cite{spark}, Obliv-C~\cite{obliv-c} and Sharemind~\cite{sharemind}
   (\S\ref{s:impl}).
\end{enumerate}

%
%
\noindent
We measured the performance of our prototype using microbenchmarks and
end-to-end analytics queries (\S\ref{s:eval}).
Even with minimal annotations on input relations and basic optimizations,
\sys scales queries to inputs orders of magnitude larger than existing
MPC frameworks support on their own.
When users annotate specific input columns, our new hybrid MPC--cleartext
protocols speed up join and aggregation operators by $7\times$ or more
compared to execution in Sharemind~\cite{sharemind}, a fast, commercial
MPC framework.
Compared to SMCQL, \sys's extra optimizations help scale a medical
research query to orders of magnitude larger inputs with comparable
security guarantees.
%

%
%

\mv{(2/15) perhaps ``a part of'' \sys's approach is compatible?}

%
%
Nevertheless, our prototype has some limitations.
It defends only against passive adversaries, though \sys's approach of minimizing MPC is
compatible with stronger threat models.
While our prototype reimplements some of SMCQL's techniques for fair comparison,
it does not support all; an ideal system would combine the two systems'
techniques.
Finally, our prototype supports only the Obliv-C and Sharemind MPC frameworks,
which limit MPC steps to two or three parties, but adding support for further
frameworks requires only modest effort.
%

%% file: background.tex
\section{Motivation and background}
\label{s:bg}


%
MPC jointly executes an agreed-upon computation across several parties'
private data without a trusted party.
MPC has served purposes from detecting VAT tax fraud by analyzing
business transactions~\cite{BogdanovJSV15}, to setting sugar beet prices via
auction~\cite{mpcgoeslive}, relating graduation rates to
employment~\cite{student-taxes}, and evaluating the gender pay gap across
businesses~\cite{BestavrosLV17}.
These applications all compute on hundreds to thousands of records,
but many useful computations on large data that might benefit from MPC are
currently infeasible.

\mv{(2/15) changed "over a few hundreds to thousands of records" since I found that awkward}

\begin{figure*}
 \centering
 \begin{subfigure}[b]{0.33\textwidth}
   \centering
   \setlength{\abovecaptionskip}{-6pt}
   \setlength{\belowcaptionskip}{0pt}
   \includegraphics{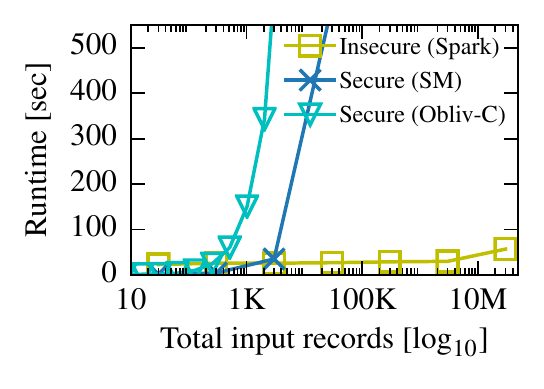}
   \caption{Aggregation (\code{SUM}).}
   \label{f:bg-motivation-agg}
 \end{subfigure}
 \begin{subfigure}[b]{0.33\textwidth}
   \centering
   \setlength{\abovecaptionskip}{-6pt}
   \setlength{\belowcaptionskip}{0pt}
   \includegraphics{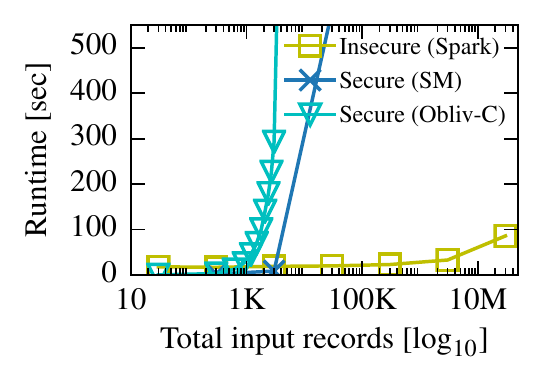}
   \caption{\code{JOIN}.}
   \label{f:bg-motivation-join}
 \end{subfigure}
 \begin{subfigure}[b]{0.33\textwidth}
   \centering
   \setlength{\abovecaptionskip}{-6pt}
   \setlength{\belowcaptionskip}{0pt}
   \includegraphics{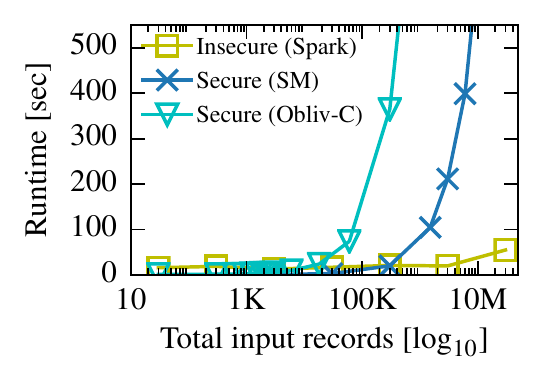}
   \caption{\code{PROJECT}.}
   \label{f:bg-motivation-project}
 \end{subfigure}
 \caption{Existing MPC frameworks only scale to small data sets for common
   relational operators, \eg aggregations and joins.
   By contrast, Spark runs these operators on tens of millions of records
   in seconds (note the log-scale $x$-axis).}
 \label{f:bg-motivation}
\end{figure*}

\subsection{Use cases for MPC on ``big data''}
\label{s:bg-examples}

We sketch two example applications of MPC that would be worthwhile if
MPC could run efficiently on large data.
\para{Credit card regulation.}
A government regulator (in the U.S., the OCC~\cite{occ-regulator}) who
oversees consumer credit reporting agencies (in the U.S., \eg TransUnion) may
wish to estimate the average credit score by geographic area (\eg ZIP code).
%
The government regulator holds the social security numbers (SSNs) and census ZIP
code of potential card holders; credit reporting agencies, by contrast, have the
SSNs of card holders, their credit lines, and their credit ratings.
By law, the government regulator cannot share the residence information.
Likewise, credit reporting agencies cannot share raw portfolios for
fear of leaking information to competitors through carelessness or compromise,
so MPC is needed.
The input to this query is large: there are over 450M SSNs~\cite{number-of-ssns}
and 167M credit cards~\cite{number-of-credit-cards} currently issued in the U.S.
%

%
\para{Market concentration.}
Competition law requires governments to regulate markets to prevent oligopolies
or monopolies.
Regulators often use the Herfindahl-Hirschman Index (HHI)\,---\,the sum of
squared market shares of companies active in a marketplace\,---\,to decide
whether scrutiny is warranted~\cite[\S5.3]{doj-hmg}.
Public revenue data is coarse-grained, and the market shares of privately-held
companies are difficult to obtain.
%
For example, airport transfers in New York City constitute a marketplace,
for which an effective HHI considers \emph{only} the revenue derived from
airport transfers in the market shares.
Airport transfers made up 3.5\% of 175M annual NYC yellow cab trips in 2014;
many trips were serviced by other vehicle-for-hire (VFH)
companies~\cite{nyc-taxi-factbook}.
%
%
While the HHI computation inputs are small (a single number per company),
computing them requires filtering and aggregating over millions of trip records
that companies keep private.
%

\subsection{Security guarantees}
\label{s:bg-guarantees}

MPC guarantees the \emph{privacy} of a computation's input and intermediate
data.
Specifically, MPC reveals no more about each party's input data than can be
inferred from the final, publicly revealed output of the computation.
MPC also guarantees \emph{correctness} of the output revealed to each party,
and can provide \emph{integrity} properties~\cite{lindell2009secure}.
%

%
MPC provides these guarantees under a specific model of its adversary (\ie
dishonest participant).
Commonly, the adversary is presumed to be \emph{passive},
respecting the protocol but trying to learn other participants' data
(the ``semi-honest'' model), or
\emph{actively malicious}, deviating from the protocol's rules to
expose private data or compromise the integrity of outputs.
Generic techniques for information-theoretically secure MPC require an honest majority~\cite{bgw, RabinB89} whereas techniques that leverage computational assumptions only require a single honest party (an ``anytrust'' model)~\cite{yao-garbled-circuits, gmw}.
%
Relaxing security guarantees makes MPC faster: honest (super-)majority 
techniques perform better than anytrust
ones~\cite{ArakiFLNO16,FurukawaLNW17,sharemind}, and a passive adversary can be withstood with
at least $7\times$ lower overhead than an actively malicious
adversary~\cite{billion-gates-mpc}.
%

\subsection{MPC techniques and scalability}
\label{sub:mpc-techniques}
\label{s:bg-scaling}

Two MPC techniques dominate today: garbled circuits and secret sharing.
Cryptographers have made efforts to scale these to many
\emph{parties}~\cite{DaniKMS14, BoyleCP15, ChandranCGGOZ15, DaniKMSZ17},
but scalability to \emph{large data} remains a challenge.
In this work, we focus on scaling to large data sizes, and assume a fixed,
small number of parties.
In \textbf{garbled circuits}~\cite{yao-garbled-circuits}, one party encrypts
each input bit
to create a ``wire label''.
Then, it converts the computation into a circuit of binary gates, each expressed
as a ``garbled truth table'' comprising a few ciphertexts.
A single evaluator party receives the circuit and the encrypted wire labels for
all input bits.
At each gate, the evaluator combines the inputs to produce an encrypted output
using the garbled truth table.
Garbled circuits require no communication between parties during evaluation, but
their state is far larger than the input data (\eg 80--128$\times$ for
typical security parameters), which makes processing of large data impractical.
\textbf{Secret sharing}~\cite{shamir-secret-sharing}, by contrast, splits each
sensitive input (\ie each integer, rather than each bit) into ``secret shares'',
which in combination yield the original data.
In a common encoding, secret shares cancel out into the cleartext value when
added together.
Each computing party works on one secret share; additions happen locally and
without communication, but multiplications require interaction before or during
the evaluation~\cite{bgw, Beaver91a, Beaver95}.
Secret sharing only multiplies state size by the number of shares, but the
communication (\ie network I/O) during computation limits scalability.
Batching communication helps reduce overheads, but a multiplication still
requires sending at least one bit between parties~\cite{ArakiFLNO16}.
%
%

%
\para{Scalability in practice.}
Figure~\ref{f:bg-motivation} compares the performance of three relational
operators in MPC frameworks based on secret-sharing
(Sharemind~\cite{sharemind}) and garbled-circuits
(Obliv-C~\cite{obliv-c}) to insecure plaintext execution.
Each experiment inputs random integers and runs a single operator.
The MPC frameworks run with two (Obliv-C) or three (Sharemind) parties,
who in aggregate contribute the record count on the log-scale $x$-axis; insecure
computation runs a single Spark job on the combined inputs.
MPC scales poorly for aggregations (Figure~\ref{f:bg-motivation-agg}) and
joins (Figure~\ref{f:bg-motivation-join}).
These operators require communication in secret sharing, and Obliv-C's state
grows fast (\eg the join runs out of memory at 30k records).
Even a projection (Figure~\ref{f:bg-motivation-project}), which requires only a
single pass over the input and no communication, fails to scale in practice.
Again, Obliv-C's circuit state growth limits its performance (it runs out of
memory at 300k records), and Sharemind takes over 10 minutes beyond
3M input records ($\approx$37 MB) due to overheads of secret-sharing and in
its storage layer.
Thus, MPC in practice only scales to a few thousand input records.
These results are consistent with prior studies: Sharemind takes 200s to sort
16,000 elements~\cite{mpc-sorting}, and DJoin takes an hour to join 15,000
records~\cite{djoin}.
Current MPC systems therefore seem unlikely to scale to even moderate-sized
data sets.
In particular, the poor performance of joins and aggregations is concerning:
more than 60\% of privacy-sensitive analytics queries use joins, and over
34\% contain aggregations~\cite[\S2]{practical-dp-for-sql}.
Short of new cryptographic techniques, the only way run MPCs on large data
may therefore be to \emph{avoid} using its cryptographic techniques unless
absolutely necessary.
%

%% file: paradigm.tex
\section{\sys overview}
\label{s:paradigm}

\sys builds on the insight that the end-to-end security guarantees of MPC
often hold even if parts of a query run outside MPC.
Intuitively, for example, any operation computed using only a party's local
inputs and publicly available data can run outside MPC, as can any operation
that applies only reversible operations and reveals their result.
\sys's guiding principle is \emph{to do as little as possible and as much as
necessary under MPC}: in other words, \sys minimizes the computation under MPC
until no further reduction is possible.
This helps scale MPC to large data by using cheaper algorithms, local
computation, and data-parallel processing systems for crucial parts of the
query.
%

%
%

\subsection{Threat model}

Like many practical MPC systems, \sys focuses on withstanding a passive,
semi-honest adversary.
The adversary can observe all network communication during execution of the
protocol, and can also statically (\ie before the protocol begins)
choose to compromise some parties to view data stored in their local file system
and memory.
Because the adversary can monitor the computation's control flow and memory access
patterns, the MPC must use \emph{oblivious} operations that avoid
data-dependent control flow and hide access patterns.
Yet, all parties\,---\,including compromised ones\,---\,faithfully execute the
MPC protocol and submit valid input data.
\extendedtr{
In the following, we restrict our attention to semi-honest security.
Appendix \ref{ss:malicious} sketches how \sys could be extended to provide
security against malicious adversaries.
}{
In this paper, we restrict our attention to semi-honest security.
In \fullversion~\cite{conclave-tr}, we also sketch how \sys can be extended to
provide security against malicious adversaries.
}
%

\subsection{Security guarantees and assumptions}
\label{s:paradigm-privacy}
\sys is a query compiler that generates code for execution in external MPC systems
(``backends'').
Consequently, \sys inherits the security guarantees and assumptions of the MPC
backend used, but must also uphold them.
In particular, \sys hides all private data processed, along with meta-data such
as value frequencies.
\sys provides these guarantees against the same threshold of colluding
adversarial parties that its MPC backend can tolerate.
%
%

%
Consistent with MPC literature, \sys treats the sizes of all \emph{input
relations} as public, and hides the sizes of intermediate relations
processed under MPC.
However, after \sys rewrites a query to move operations out of MPC, the
sizes of inputs to the remaining MPC may differ from the original input
sizes.
Depending on the query, this may leak information.
\sys's rewrites are safe if the sizes of the new MPC input relations are
data-independent. 
For data-dependent sizes, \sys only proceeds with a rewrite if all affected
parties have authorized it, choosing a slower query plan otherwise.
Additionally, \sys provides \emph{hybrid protocols} that provide parties with
the option to trade some security for better performance.
%
These hybrid protocols outsource work for cleartext processing at a chosen
party, and selectively reveal some data to it. 
\sys applies such value-leaking rewrites \emph{only} if parties supply explicit
input annotations and \sys can derive an authorization. 
Hybrid protocols essentially create a ``server-aided'' setting
\cite{DBLP:journals/iacr/KamaraMR11,DBLP:conf/ccs/KamaraMR12} with leakage:
the aiding \emph{selectively-trusted party} (STP) catalyzes the MPC by performing
otherwise expensive operations outside of MPC's cryptographic guarantees.
%
%
Only a single STP can exist in a \sys execution; all other participants are
\emph{regular parties}.
\nv{(2/15) Added below clarification, please review.}
Which party takes on the role of an STP depends on the specific trust assumptions of a given deployment; one of the data contributors may act as an STP, but it may also be a party without inputs whose sole role is to assist the MPC.
%
%
%
%
%
\sys retains MPC's security guarantees against any adversary who compromises the
STP alone or a subset of regular parties that the MPC backend can withstand.
\sys makes no guarantees against an adversary who compromises both regular parties and the STP.


%% file: system.tex
\section{Specifying \sys queries}
\label{s:specification}

\begin{lstlisting}[language=conclave,
  caption={Credit card regulation query in \sys's
           LINQ-style frontend with input relations
           locations (lines 6, 10--11),
           and an optional trust annotation (line 8).},
           %on the
           %\code{ssn} input column of party B's and C's
           %input tables (line\ 8).},
  label=l:ssn-workflow, float, floatplacement=t]
import conclave as cc
pA, pB, pC = cc.Party("mpc.ftc.gov"), \
      cc.Party("mpc.a.com"), cc.Party("mpc.b.cash")
demo_schema = [Column("ssn", cc.INT, trust=[]),
                  Column("zip", cc.INT, trust=[])]
demographics = cc.newTable(demo_schema, at=pA)
# banks trust the regulator to compute on SSNs
bank_schema = [Column("ssn", cc.INT, trust=[pA]),
                  Column("score", cc.INT, trust=[])]
scores1 = cc.newTable(bank_schema, at=pB)
scores2 = cc.newTable(bank_schema, at=pC)
scores = cc.concat([scores1, scores2])
# query to compute average credit score by ZIP
joined = demographics.join(scores, left=["ssn"],
                               right=["ssn"])
by_zip = joined.aggregate("count", cc.COUNT,
                              group=["zip"])
total_sc = joined.aggregate("total", cc.SUM,
                                 group=["zip"])
avg_scores = \
  total_sc.join(by_zip, left=["zip"], right=["zip"])
            .divide("avg_score", "total", by="count")
# regulator gets the average credit score by ZIP
avg_scores.writeToCSV(to=[pA])
\end{lstlisting}

\sys compiles a relational query into executable code.
It is similar to plaintext-only big data query compilers (\eg Hive~\cite{hive},
Pig~\cite{pig}, or Scope~\cite{scope}) and ``workflow managers'', like
Apache Oozie~\cite{oozie} or Musketeer~\cite{musketeer}.
Like these systems, \sys transforms the query into a directed acyclic graph
(DAG) of relational operators, and it executes this DAG on one or more
backend systems.
Unlike prior query compilers, \sys's query rewrite rules must preserve
MPC's security guarantees while aiming to improve performance.
%

\subsection{Assumptions}
\label{s:design-assumptions}


%
\sys assumes that analysts write relational queries using SQL or
LINQ~\cite{linq}, that parties agree via out-of-band mechanisms on the
query to run, and that all parties faithfully execute the protocol.
Parties locally store input data with the schema expected by the
query.
Each party runs \one{} a local \sys agent that communicates with the other
parties and manages local and MPC jobs, \two{} a local MPC endpoint (\eg an
Obliv-C or Sharemind node), all on private infrastructure and \three{}
optionally, a parallel data processing system (\eg Spark).
Absent a parallel data processing system, \sys runs local computation in
sequential Python.

\subsection{Query specification}
\label{s:workflow-specification}

\begin{lstlisting}[language=conclave,
  caption={Market concentration query in \sys's LINQ-style frontend. Note
           the owner annotations on the input tables (lines 8--10) and the
           final result (line 27).}, label=l:hhi-workflow, float, floatplacement=t]
import conclave as cc
pA, pB, pC = cc.Party("mpc.a.com"), \
       cc.Party("mpc.b.com"), cc.Party("mpc.c.org")
# 3 parties each contribute inputs with same schema
schema = [Column("companyID", cc.INT, trust=[]),
            # ...
            Column("price", cc.INT, trust=[])]
inputA = cc.newTable(schema, at=pA)
inputB = cc.newTable(schema, at=pB)
inputC = cc.newTable(schema, at=pC)
# create multi-party input relation
taxi_data = cc.concat([inputA, inputB, inputC])
# relational query
rev = taxi_data.project(["companyID", "price"])
        .aggregate("local_rev", cc.SUM,
                     group=["companyID"], over="price")
        .project([0, "local_rev"])
market_size = rev.aggregate("total_rev", cc.SUM,
                                 over="local_rev")
share = rev.join(market_size, left=["companyID"],
                    right=["companyID"])
             .divide("m_share", "local_rev",
                      by="total_rev")
hhi = share.multiply(share, "ms_squared", "m_share")
           .aggregate("hhi", cc.SUM, on="ms_squared")
# finally, party A gets the resulting HHI value
hhi.writeToCSV(to=[pA])
\end{lstlisting}

\sys queries can be written in any way that compiles to a directed acyclic
graph (DAG) of operators.
Listings~\ref{l:ssn-workflow} and \ref{l:hhi-workflow} show the credit
ratings and market concentration queries from \S\ref{s:bg-examples} in a
DryadLINQ-like language~\cite{dryadlinq}.
Even though the input data to a \sys query is distributed across multiple
parties, \sys largely abstracts this fact away from analysts.
Instead, the analysts specify the query's core as though it was a relational
query on a single database stored at a trusted party
(lines 14--22 of Listing~\ref{l:ssn-workflow} and
lines 14--25 of Listing~\ref{l:hhi-workflow}).
The only difference to a simple query is that each input relation has an
``owner'', \viz the party storing it, supplied via an \code{at} annotation
(lines 6--11 and 5--10).
This information helps \sys{} both locate the data and derive where operations
combine data across parties.
By combining per-party input relations using a duplicate-preserving set union
operator (\code{concat}, lines 12 and 12), analysts can create compound
relations across parties and use them in the query.
In addition to inputs, analysts also annotate each output relation with one or
more recipient parties (\code{to}).
These parties receive the cleartext result of executing the query
(lines 24 and 27).
\sys's high-level, declarative query specification contrasts with existing
MPC frameworks, which usually provide Turing-complete languages (\eg
SecreC~\cite{secrec} or Obliv-C~\cite{obliv-c}).
Such interfaces are expressive, but are often unfamiliar to data analysts
and require fine-grained security annotations of intermediate variables.
%

%
%
%
%

\begin{figure*}
 \centering
 \includegraphics[width=\textwidth]{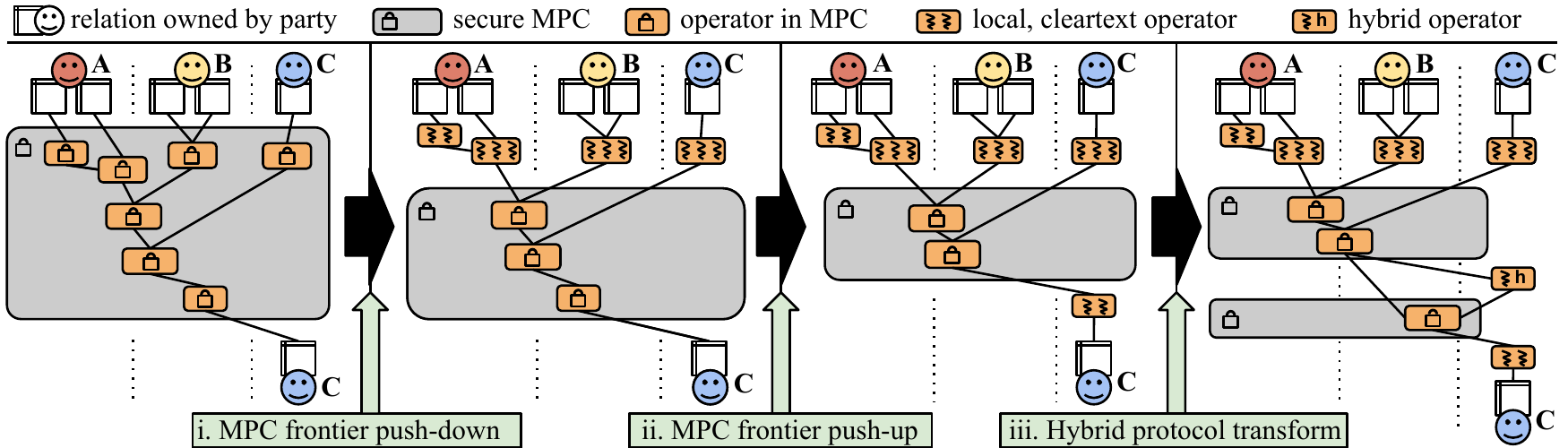}
 \caption{\sys minimizes the work under MPC by:
          \one{} pushing the MPC frontier down and locally preprocessing where
          possible; \two{} pushing the MPC frontier up from the outputs,
          processing reversible operators in the clear at the receiving party;
          and \three{} inserting special ``hybrid'' operators that implement
          efficient hybrid MPC-cleartext protocols. In this example, the
          rightmost party (C, blue) contributes data and also acts as a
          selectively-trusted party for the hybrid join operator.}
 \label{f:analysis-rewrite-pipeline}
\end{figure*}


\subsection{Optional trust annotations}
\label{s:design-annotations}

In addition to mandatory input locations, \sys also supports optional,
light-weight \emph{trust annotations} that help it apply further optimizations.
These annotations specify parties who are authorized to learn values
in specific input schema columns in the clear to compute more efficiently on
them.
%
%
%

%
The intuition behind trust annotations is that the sensitivity of data within a
relation often varies by column.
Consider a relation that holds information about a company's branches: it may
have public \code{address} and \code{zip} columns (as this information is
readily available from public sources), but privately-owned columns
\code{manager} or \code{turnover}.
Other columns may be private, but the owning party might be happy to reveal them
to a specific \emph{selectively-trusted party} (STP), such as a government
regulator.
For example, in the credit card regulation query
(Listing~\ref{l:ssn-workflow}), the government regulator already holds
demographic information organized by SSN, and the credit card companies may be
willing to reveal the SSNs of their customers to the regulator (though not to
the other parties, who are competitors).
Hence, the parties agree to make the regulator an STP for the
\code{ssn} column of the credit card companies' customer relations (see
Listing~\ref{l:ssn-workflow}, line 8).
Such selective revealing of columns helps \sys avoid, or shrink, expensive
MPC steps and substantially improves performance.
A trust annotation associates a column definition with a \emph{trust set} of
one or more parties.
Any party in the trust set can be an STP for computing on the annotated column.
This party may obtain the cleartext data for this column and combine it%
\,---\,locally and in the clear\,---\,with public columns, other columns with
an overlapping trust set, and columns it privately owns.
A party storing an input relation is implicitly in the trust set for all its
columns, as are recipients of an output relation.
Finally, a public column has all parties in its trust set.
%

%
%

\section{Query compilation}
\label{s:compilation}

The annotations on input and output relations enable \sys to determine which parts of the query DAG must run under MPC.
\sys automates this reasoning to free data analysts from manual labor and
to avoid subtle mistakes.
Its goal is to execute as many operators as possible outside of MPC, and to
reduce data volume processed under MPC where possible, while maintaining MPC's
security guarantees.
To achieve this, \sys applies a combination of static analysis, query rewriting
transformations, and partitioning heuristics.
\sys compiles a query in six stages
(partly illustrated in Figure~\ref{f:analysis-rewrite-pipeline}); all parties run
these stages deterministically.
\begin{enumerate}[nosep]
  \item \sys starts with a query plan consisting of a single, large MPC.
    First, it propagates input relation locations to intermediate relations
    to determine where data crosses party
    boundaries (\S\ref{s:design-propagation}).
  \item Using this information, \sys then rewrites the query into an equivalent
    query with fewer operators under MPC.
    This results in a DAG with a clique of inner operators under MPC, and with
    efficient cleartext operators at the roots and leaves
    (\S\ref{s:design-pushupdown}).
  \item \sys then propagates the trust annotations from input relations through
    the DAG, and combines them according to inference rules in order
    to determine when parts of operators can run outside MPC.
  \item Subsequently, and propagated trust annotations permitting, \sys splits
    the monolithic inner MPC into several smaller MPCs and local steps by adding
    hybrid protocol operators in place of operators that can run partially
    outside MPC (\S\ref{s:design-hybrid}).
  \item \sys further minimizes the use of expensive oblivious sub-protocols, such
    as sorts, by moving these operations into local processing or replacing them
    with cheaper equivalents if possible (\S\ref{s:design-min-obliv}).
  \item Finally, \sys partitions the query by splitting the DAG at each
    transition between local and MPC operations, generates code for the
    resulting sub-DAGs, and executes them on the respective backends.
\end{enumerate}
Note that all transformations that \sys applies do improve end-to-end query
runtime, but they do \emph{not} strictly reduce the overall work.
In fact, some transformations create \emph{additional} local processing work for parties,
or they reduce the efficiency of local cleartext processing in exchange for doing
less work under MPC.
For example, a join between an intermediate relation and a public relation (\eg
a relation mapping ZIP codes to geolocations) might process fewer records if
it runs late in the query (\eg after filters and aggregations).
Conventional query optimizers would apply filters before joins, but this may
force the join into MPC, which is much slower than a local join of the
input data against the public relation.
Hence, doing the join several times (once per party) and against more rows
actually speeds up the query.
%

\subsection{Propagating annotations}
\label{s:design-propagation}

The input and output relation annotations give \sys information about the roots
and leaves of the DAG.
\sys propagates this information through the DAG in two passes, which infer the
execution constraints on its operators.
The \textbf{first pass} propagates input locations down the DAG in a
topological order traversal, and propagates output locations back up the graph
in a reverse topological order traversal.
For each intermediate operator, the propagation derives the \emph{owner} of its
result relation.
A party ``owns'' a relation if it can derive it locally given only its own data.
Input and output relations are owned by the parties that store them.
Relation ownership propagates along edges according to inference rules.
The output of a unary (\ie single-input) operator inherits the ownership of
its input relation directly.
The owner of the output relation of a multi-input operator depends on
ownership of its input relations.
If all input relations have the same owner, that owner propagates to the
output relation; if they have different owners, the output relation has no
owner.
%
%
This process captures the fact that no single party can compute output that
combines different parties' data.
Operators with output relations that lack an owner \emph{must} run
under MPC.

%


%
In a \textbf{second pass}, \sys determines the trust set (cf. \S\ref{s:design-annotations}) for each intermediate relation's columns.
A party is ``trusted'' with an intermediate result column if it is entrusted with
enough input data in order to calculate that column in the clear.
Propagating this notion of trust enables \sys to use hybrid protocols to compute
intermediate relations, while ensuring that no more information is leaked than
was explicitly authorized via the parties' trust annotations on their input schema.
Columns within the same relation may have different trust sets, because knowing one
column of a relation does not imply the ability to calculate any other columns.
%

%
The trust set for each input column is defined by the trust annotation.
\sys propagates these trust annotations down the DAG in topological order.
For each result column $c$ of each operator $o$, \sys determines which of $o$'s
operand columns contribute toward the calculation of $c$.
\sys sets the trust set of $c$ as the intersection of trust sets of these operand columns.
The dependencies between operand and result columns are defined by operator semantics
in two ways.
First, a result column depends on all operand columns that directly contribute rows to it.
For example, the first result column of a \code{concat} depends on the first column of each
concatenated relation, since its rows are derived from those operand columns.
Second, a result column depends on all operand columns that affect how its rows are
combined, filtered, or reordered.
For example, the computed column of a \code{sum} depends both on the column aggregated over
\emph{and} the group-by columns, since the group-by columns determine how the aggregated
column rows are combined.
Likewise, all result columns of a \code{join} depend on the join key columns, which
determine whether a row is part of the output.
\sys encodes column dependencies for all operator types it supports.
This propagation algorithm helps \sys maintain an important security invariant: \sys only
reveals a column to a party if the column can be derived from input columns that the
party is authorized to learn.
We use this guarantee to ensure that \sys's use of hybrid protocols is safe.
\subsection{Finding the MPC frontier}
\label{s:design-pushupdown}

\sys starts planning the query with the entire DAG in a single, large MPC.
It then pulls operators that can run on local cleartext data out of MPC, and
splits other operators into local preprocessing operators and a smaller MPC
step.
These transformations push the \emph{MPC frontier}\,---\,\ie the boundary
between MPC operators and local cleartext operators\,---\,deeper into the DAG,
where a clique of operators remains under MPC. The transformations have little to no impact on the cryptographic security guaranteed by \sys's backend.

\para{MPC frontier push-down.}
Starting from the input relations,
\sys pushes the MPC frontier down the DAG as far as possible while preserving
correctness and security.
After the ownership propagation pass, each relation is either \one{} a
\emph{singleton relation} with a unique owner; or \two{} a
\emph{partitioned relation} without an owner.
In a partitioned relation, multiple parties hold a subset (\ie partition) of the
relation.
\sys traverses the DAG from each singleton input relation and pulls operators
out of MPC until it encounters an operator with a partitioned output,
which it must process under MPC.
Queries often combine inputs from multiple parties into a single, partitioned
relation via a \code{concat} operator.
This creates a ``virtual'' input relation that contains data from all
parties (\eg \code{scores} on line 12 of Listing~\ref{l:ssn-workflow},
and \code{taxi\_data} on line 12 of Listing~\ref{l:hhi-workflow}).
While convenient, this forces \sys to enter MPC early, since the output of
a \code{concat} operator is a partitioned relation.
To avoid this, \sys pushes \code{concat} operators down past any operators
that are distributive over input partitions: \ie for operator \code{op} and
relations $R_{pA}$ to $R_{pN}$ owned by \textit{pA} to \textit{pN},
\[
\texttt{op}(R_{pA}~|~...~|~R_{pN}) \equiv \texttt{op}(R_{pA})~|~...~|~\texttt{op}(R_{pN}).
\]
For example, the projection over \code{taxi\_data} in the market
concentration query (Listing~\ref{l:hhi-workflow}, line 13) is distributive,
as applying it locally to \code{inputA}, \code{inputB}, and \code{inputC}
produces the same result and leaks no more information than running it under
MPC.
Consequently, \sys can push the MPC frontier further down.
Other operators, however, do not trivially distribute over the inputs of a
\code{concat}.
For example, to split an aggregation that groups a partitioned relation by
key, \sys adds per-party aggregations over singleton relations, followed by
a secondary aggregation.
This transformation may have security implications, which we discuss at the
end of this section.
%
%
While the secondary aggregation must remain under MPC, \sys can pull the
local preaggregations out of MPC and significantly reduce the MPC's input
data size.
Moreover, \sys can push the operator clique of \code{concat} and the
secondary aggregation (which now forms the MPC frontier) past other
distributive operators.
In the market concentration query, \sys pushes the MPC frontier to right
above the \code{market\_size} relation (Line 17, Listing~\ref{l:hhi-workflow}), at which point the data amount to
only few integers per party.
\para{MPC frontier push-up.}
In certain cases, \sys can also push the MPC boundary up from the output
relations (\ie the DAG's leaves).
Some relational operators are \emph{reversible}, \ie given their output, it
is possible to reconstruct the input without additional information.
For example, multiplication of column values by a fixed non-zero scalar has
this property.
%
For a reversible operator, the output of the operation fundamentally \emph{leaks} its input, and hence it
need not run under MPC.
Instead, \sys reveals the reversible operator's input relation to the final
recipients for local cleartext multiplication, even if the input has a
different owner.
\sys's MPC push-up pass starts at output relations and lifts the MPC
frontier through reversible operators.
Key examples are arithmetic operations and reordering projections; additionally,
while aggregations are generally not reversible, special cases are.
If \code{count} occurs as a leaf operator, then it inherently reveals the
group-by key frequencies, and \sys can rewrite it to an MPC projection
and a cleartext count.
The projection removes all columns apart from the group-by columns (under
MPC), and the recipients count the keys in the clear.
As projections are more scalable under MPC than aggregations
(\S\ref{s:bg-scaling}), this improves performance.

\para{Security implications.}
\sys's push-down and push-up query transformations can affect security
guarantees.
Without any transformations (\ie if the entire query remains within the MPC
frontier), \sys directly inherits the security guarantees provided by its MPC
backend.
\sys's push-up operations have no impact on security because they are
reversible, and thus the input to the operator is simulatable from its output.
However, push-downs do have a security implication: they change the lengths of
inputs to the MPC backend.
For instance, splitting an aggregation into a local step and an MPC step (even
as part of a ``secondary aggregation'' as described above) results in the
parties learning the count of distinct group-by column values contributed by
each party, as opposed to the total number of records per party.
For this reason, \sys requires the consent of parties to make any push-down
transformations that result in data-dependent cardinalities at the input to
secure MPC.
In the appendix, we formally prove that the leaked input length information
is the only security implication of \sys's MPC frontier changes (see
Theorem~\ref{thm:composition-without-hybrids}).

\subsection{Hybrid operators}
\label{s:design-hybrid}

In this rewrite pass, \sys splits work-intensive operators, such as joins
and aggregations, into \emph{hybrid operators}.
Hybrid operators outsource expensive portions of an operator to a
selectively-trusted party (STP) by revealing some input columns to the STP.
Hybrid operator execution thus involves local computation at the STP and
MPC steps across all parties.
\sys \emph{only} applies this transformation if the query's trust
annotations relax input columns' privacy constraints.
%

%
%
In the context of hybrid operators, a party is either the STP, or
a regular, untrusted party.
\sys exposes the plaintext \emph{values} of some columns to the
STP, and the size of the result (\ie row count) to all parties; otherwise,
it maintains full MPC guarantees for these columns towards the untrusted
parties, and towards all parties for all other columns.

\sys currently supports three hybrid operators: a \emph{hybrid join}, a
\emph{public join}, and a \emph{hybrid aggregation}.

\begin{figure*}
  \begin{tikzpicture}
    \node[anchor=west, align=left] at (-1.1, 0) {\lstinline[language=conclave]{def hybridJoin(left: Relation, right: Relation, left_key_col: Column, right_key_col: Column, STP: Party):}};
    \filldraw[fill=gray!40!white] (-1, -0.25) rectangle (5.95, -5.05);
    \node[anchor=north east, align=right] at (5.95, -0.25) {\large \bf MPC};
    \node[anchor=east] at (-0.5, -0.5) {\lstinline{1}};
    \node[anchor=west, align=left] at (-0.5, -0.5) {\withlineno{1}{left.shuffle()}};
    \node[anchor=east] at (-0.5, -0.8) {\lstinline{2}};
    \node[anchor=west, align=left] at (-0.5, -0.8) {\withlineno{2}{right.shuffle()}};
    \node[anchor=east] at (-0.5, -1.1) {\lstinline{3}};
    \node[anchor=west, align=left] (lk) at (-0.5, -1.1) {\withlineno{3}{left_keys =  left.project([left_key_col])}};
    \draw [->] (0, -1.25) -- ++(0, -0.15) -- node[pos=0.8, above=0pt] {\code{revealTo(STP)}} ++(9, 0);
    \node[anchor=east] at (-0.5, -1.65) {\lstinline{5}};
    \node[anchor=west, align=left] (rk) at (-0.5, -1.65) {\withlineno{5}{right_keys =  right.project([right_key_col])}};
    \draw [->] (0, -1.8) -- ++(0, -0.15) -- node[pos=0.8, above=0pt] {\code{revealTo(STP)}} ++(9, 0);
    \node[anchor=north west, align=left] at (9, -0.7) {\large \textbf{STP} (in cleartext)};
    \node[anchor=east] at (9.4, -1.35) {\lstinline{4}};
    \node[anchor=west, align=left] at (9.4, -1.35) {\withlineno{4}{left.enumerate()  # [a, 0], [b, 1], ...}};
    \node[anchor=east] at (9.4, -1.9) {\lstinline{6}};
    \node[anchor=west, align=left] at (9.4, -1.9) {\withlineno{6}{right.enumerate()  # [c, 0], [a, 1], ...}};
    \node[anchor=east] at (9.4, -2.2) {\lstinline{7}};
    \node[anchor=west, align=left] at (9.4, -2.2) {\withlineno{7}{joined = left.join(right)  # [a, 0, 1]}};
    \node[anchor=east] at (9.4, -2.5) {\lstinline{8}};
    \node[anchor=west, align=left] at (9.4, -2.5) {\withlineno{8}{joined.project([left_idx]) # [0], ...}};
    \draw [->] (11.0, -2.65) -- ++(0, -0.15) -- node[above=0pt, pos=0.4] {\code{inputToMPC}} ++(-9.35, 0);
    \node[anchor=east] at (9.4, -3.05) {\lstinline{10}};
    \node[anchor=west, align=left] at (9.4, -3.05) {\withlineno{10}{joined.project([right_idx]) # [1], ...}};
    \draw [->] (11.0, -3.15) -- ++(0, -0.15) -- node[above=0pt, pos=0.4] {\code{inputToMPC}} ++(-9.35, 0);
    \node[anchor=east] at (-0.5, -2.75) {\lstinline{9}};
    \node[anchor=west, align=left] at (-0.5, -2.75) {\withlineno{9}{left_indexes}};
    \node[anchor=east] at (-0.5, -3.30) {\lstinline{11}};
    \node[anchor=west, align=left] at (-0.5, -3.30) {\withlineno{11}{right_indexes}};
    \node[anchor=east] at (-0.5, -3.60) {\lstinline{12}};
    \node[anchor=west, align=left] at (-0.5, -3.60) {\withlineno{12}{left_rows = left.select(left_indexes)}};
    \node[anchor=east] at (-0.5, -3.90) {\lstinline{13}};
    \node[anchor=west, align=left] at (-0.5, -3.90) {\withlineno{13}{right_rows = right.select(right_indexes)}};
    \node[anchor=east] at (-0.5, -4.20) {\lstinline{14}};
    \node[anchor=west, align=left] at (-0.5, -4.20) {\withlineno{14}{for (l, r) in (left_rows, right_rows):}};
    \node[anchor=east] at (-0.5, -4.50) {\lstinline{15}};
    \node[anchor=west, align=left] at (-0.2, -4.50) {\withlineno{15}{joined.append(l.concat(r))}};
    \node[anchor=east] at (-0.5, -4.80) {\lstinline{16}};
    \node[anchor=west, align=left] at (-0.5, -4.80) {\withlineno{16}{return joined.shuffle()}};
  \end{tikzpicture}
  \caption{For the hybrid join, \sys augments MPC with local computation to speed up execution.
    The protocol performs oblivious shuffles (lines 1--2) followed by a cleartext enumeration, a join, and
    a projection (lines 4--10) outside MPC (at the STP), and finishes with
    inexpensive oblivious indexing to reconstruct the join result under MPC (lines 12--16).
    The \code{revealTo} and \code{inputToMPC} operations move data in and out of MPC;
    \code{revealTo} reveals secret data to a specific party, while \code{inputToMPC} inputs a
    local dataset into MPC, for instance via secret-sharing.}
  \label{f:design-hybrid-join}
\end{figure*}

\para{Hybrid join.}
\sys can transform a regular MPC join into a hybrid join if the key columns of
\emph{both} sides of the join have intersecting trust sets, \ie they share an
STP.
This STP learns the key columns on \emph{both} sides of the join
and computes, in the clear, which keys match.
The STP hence determines which rows in the secret-shared relations are in
the join result without learning any other column values.
%

%
The protocol proceeds as follows (Figure~\ref{f:design-hybrid-join}):
\begin{enumerate}[nosep]
 \item The parties obliviously shuffle the input relations under MPC (lines 1--2 of Figure~\ref{f:design-hybrid-join}).
 %
 \item \label{enum:hybrid-join-stp} The parties project away all columns other than the
   join key columns, and reveal the resulting key-column-only relations to
   the STP (lines 3 and 5).
   %
 %
 \item For each key-column relation, the STP enumerates the rows (lines 4 and 6).
   The enumeration assigns a unique identifier to each input row, which \sys
   later uses to link the joined results back to rows in the
   secret-shared, shuffled input relations still protected by MPC.
 \item The STP performs a cleartext join on the enumerated
   key-column relations (line 7).
   This produces a set of rows that each contain the join key and two unique
   row identifiers for the left and right rows joined.
 \item \label{enum:Hjoin-regular-first} The STP projects the row identifier columns for the left and right relations into two new relations and secret-shares these to the untrusted parties (lines 8--11).
 \item \label{enum:hybrid-join-mpc} Back under MPC, the parties perform an oblivious indexing protocol (the \code{select} operator) akin to the one by Laud~\cite{Laud15} for each shuffled input relation (lines 12 and 13).
 Given a set of secret indexes, the indexing protocol obliviously selects the rows at the corresponding positions from a relation.
 Using the index relations and shuffled inputs, the parties select the left and right rows comprising the result, producing two relations \code{left_rows} and \code{right_rows}.
 \item \label{enum:Hjoin-regular-last} The parties  concatenate the two relations column-wise and obliviously reshuffle the result (lines 14--16).
 %
\end{enumerate}
\noindent
%
%
The hybrid join gives an asymptotic improvement over a standard MPC join: the oblivious indexing protocol in Step 6 requires
$\mathcal{O}((n + m)\log{}(n + m))$ non-linear operations, where $n$ is
the input size and $m$ the result size, whereas an
MPC join requires $\mathcal{O}(n^{2})$ non-linear
operations (but assumes no STP).
%

%
%

\para{Public join.}
\sys can use the public join operator if the join key columns of \emph{both}
sides of an MPC join are public.
This is the case if both columns' trust annotations include all parties, and
hence \emph{any} party may learn the key column values (though not other
column values).
The public join proceeds exactly as the hybrid join but without the oblivious
indexing and shuffle steps: the parties send the join key columns to a randomly selected party; the party enumerates and joins the rows, and finally sends the joined index relation to all parties, who
compute the joined result.
Even though some MPC frameworks have built-in cleartext processing capabilities,
\sys uses this approach since it allows the use of a data-parallel framework
(\eg Spark) for local work.
%



%

\para{Hybrid aggregation.}
\sys can transform an MPC aggregation into a hybrid aggregation if the trust
set on the group-by column contains an STP.
%
%


%
\sys's hybrid aggregation protocol adapts the sorting-based MPC protocol by
J\'{o}nsson \emph{et al.}~\cite{mpc-sorting}.
In the original protocol, the parties arrange the rows into key-groups by sorting on the key, obliviously accumulate the aggregate for each key-group into its last entry, and discard the other entries.
In the hybrid version, the STP can perform the sort in the clear and assist the other parties in the accumulation step.
It proceeds as follows:

\begin{enumerate}[nosep]
  \item \label{enum:hybrid-agg-stp} The parties obliviously shuffle the input and reveal the shuffled group-by column to the STP.
    We refer to the entries in the group-by column as \emph{keys}.
  \item Locally, the STP enumerates the revealed keys, producing a relation with a key column and an index column.
    The STP sorts the relation on the key column.
    This groups together all rows with equal keys, and creates a mapping for each row in the sorted relation to its original position via the indexes.
  \item The STP scans over the relation and computes an \emph{equality flag} for each row that indicates whether the keys for this row and the previous row are equal.
  %
  \item The STP projects away the key column from the sorted relation, leaving only the indexes.
    The STP sends the indexes to the other parties \emph{in the clear}.
  \item \label{enum:Hagg-regular-first} The STP secret-shares the equality flags.
  \item Using the plain-text ordering information, the untrusted parties
    reorder the rows of the shuffled relation so that the rows are sorted by
    group-by column values.
  \item Under MPC, the parties scan over the result. For each entry, they obliviously aggregate the previous value
    into the current if the corresponding secret equality flag is one.
    This accumulates the aggregate for each key group into the group's last entry.
    At each entry, the parties also store the secret equality flag for the last
    comparison; the flag is set unless the entry is the last one in the
    group.
  \item \label{enum:Hagg-regular-last} The parties shuffle the result and reveal the equality flags; they
    discard all entries with the flag set.
\end{enumerate}

\noindent
The hybrid aggregation improves asymptotically over the regular MPC protocol: the oblivious sorting step of the original protocol requires $\mathcal{O}(n\log^{2}n)$ oblivious comparisons; in contrast, the hybrid aggregation performs the sort in the clear and only needs an oblivious shuffle which can be realized with $\mathcal{O}(n\log{}n)$ multiplications.
However, the hybrid aggregation leaks the size of the result to all untrusted parties.

\para{Security implications.}
\sys's hybrid operators introduce two new types of leakage: the STP learns
authorized columns and all parties (regular and STP) observe the cardinalities
of inputs and outputs to the hybrid relation.
\sys only uses hybrid operators if it can derive appropriate authorizations
for this leakage from input column annotations.
In the appendix, we prove that, modulo this leakage, \sys continues to achieve
MPC's simulation-based security guarantee when using hybrid operators
(Theorem \ref{thm:composition-with-hybrids}).
At a high level, we prove this by partitioning the computation into three
distinct MPC stages\,---\,before, during, and after the hybrid
operator\,---\,using the fact that simulation-based security definitions
provide sequential composition.
We show that the leakage stated above makes the view of the STP simple to
simulate and has little impact on \sys's security against a coalition of
regular parties.
\subsection{Reducing oblivious operations}
\label{s:design-min-obliv}

\sys's relational operator implementations rely on MPC ``sub-protocols'' such
as oblivious sorts and shuffles (\eg the aggregation protocol by J\'{o}nsson \emph{et al.}~\cite{mpc-sorting} uses sorts).

These operations, especially sorts, are expensive under MPC, since they must
remain control-flow agnostic.
However, if an operator produces a sorted relation, subsequent sorts can in some cases be eliminated.
For instance, if a query consists of an order-by operation followed by an oblivious aggregation, it is unnecessary to sort as part of the aggregation, as the relation is already in the correct order.
\sys thus minimizes the use of oblivious sorts by traversing the DAG, tracking the columns by which intermediate relations are sorted (if any), and eliminating redundant sorts.
Some operations, such as shuffles, do not preserve row order; \sys's propagation therefore also tracks when a sorted relation is permuted and marks the result as unsorted.
This optimization yields especially high performance gains when a relation is sorted in the clear (for instance by a public column) as it allows \sys, depending on the query, to avoid oblivious sorting altogether.

\extendedtr{
The above optimization only tracks redundant sorts; we can further improve it by moving sorts up through the DAG into cleartext processing.
Sort operations can move across any order-preserving operator (such as
a filter).
If a sort reaches a relation in which the sort order column is public, \sys
can sort the rows in the clear.
Likewise, if a sort reaches a relation owned by a single party, the party can
perform the sort in the clear.
While \code{concat} operations are not order-preserving, \sys can still push
the sort through the \code{concat} by inserting after it a merge operation.
The merge takes several sorted relations and obliviously merges them, which is
cheaper than obliviously sorting the entire data.

In future work, we plan to extend \sys with these improvements.
These transformations are another example of a key idea in \sys: doing more
work locally (\eg redundantly sorting duplicate rows later eliminated) can
yield lower execution times than doing reduced work under MPC.
} {

}

%% file: impl.tex
\section{Implementation}
\label{s:impl}

We implemented \sys as a query compiler architected similarly to ``big
data'' workflow managers like Musketeer~\cite{musketeer}.
Our prototype consists of 8,000 lines of Python, and currently integrates
sequential Python and data-parallel Spark~\cite{spark} as cleartext
backends, and Sharemind~\cite{sharemind} and Obliv-C~\cite{obliv-c} as MPC
backends.
As \sys's interfaces are generic, adding other backends requires modest
implementation effort.
%


%
Our prototype supports table schema definitions, relational operators
(join, aggregate, project, filter) as well as enumeration, arithmetic on columns
and scalars.
This captures many practical queries: \sys's current query support appears,
for example, sufficient to express 88\% of 8M sensitive real-world queries at
Uber~\cite{practical-dp-for-sql}.
%

%
%
%

%
%




We implemented the same standard MPC algorithms for joins (a Cartesian
product approach) and aggregations~\cite{mpc-sorting} in both Sharemind and
Obliv-C.

%% file: eval.tex
\section{Evaluation}
\label{s:eval}

We evaluate \sys using our motivating queries (\S\ref{s:bg})
as well as microbenchmarks.
We seek to answer:
\begin{enumerate}[nosep]
%
%
 \item How does the runtime of realistic queries running on \sys scale
   as data size grows? (\S\ref{s:eval-hhi}, \S\ref{s:eval-ssn})
 \item What impact on performance do \sys's trust annotations and
   hybrid operators have? (\S\ref{s:eval-ubench})
 \item How does \sys compare to SMCQL~\cite{smcql}, a state-of-the-art
   query processor for MPC? (\S\ref{s:eval-smcql})
\end{enumerate}
\noindent
\para{Setup.}
%
Unless otherwise specified, we run our experiments with three
parties; each party runs a four-node cluster that consists of three
Spark VMs and one Sharemind VM.
The Spark VMs have 2 vCPUs (2.4 GHz) and 4 GB RAM, and run Ubuntu 14.04
with Spark 2.2 and Hadoop 2.6.
The Sharemind VM has 4 vCPUs (2.4 GHz) and 8 GB RAM, and runs Debian
Squeeze and Sharemind 2016.12.

\noindent
\para{Metrics.}
All our graphs increase the data size on the $x$-axis by five to
eight orders of magnitude, and plot query runtime on the $y$-axis.
Less is better in all graphs, and we use a $\log_{10}$-scale $x$-axis
to be able to show the scalability limits of different systems on the
same graph. 
%

\subsection{Market concentration query}
\label{s:eval-hhi}

\begin{figure}[t]
  \centering
  \setlength{\belowcaptionskip}{-2pt}
  \includegraphics{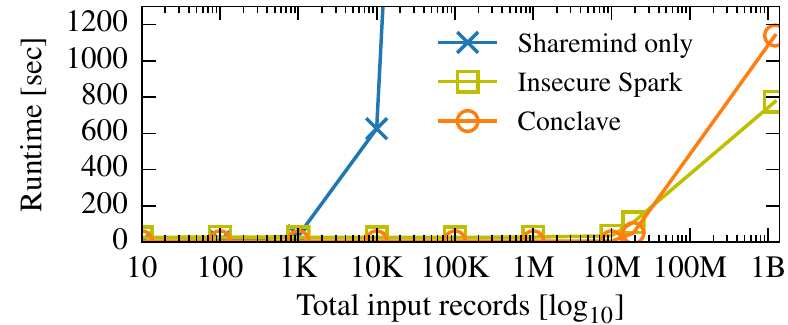}
  \caption{\sys runs the market concentration query in $<$20 minutes
           for 1B input records; under Sharemind MPC, the query cannot
           scale past 10k input records.}
  \label{f:hhi}
\end{figure}

The market concentration query computes the Herfindahl Hir\-schman Index
(HHI)~\cite{hhi} over the market shares of several vehicle-for-hire (VFH)
companies, whose sales books we model using six years of public NYC taxi
trip fare information~\cite{nyc-taxi-trips}.
%
%
We randomly divide the trips across three imaginary VFH companies and filter
out any trips with a zero fare, resulting in a total 1.3 billion trips across
all parties.
We subsample different numbers of rows from the input data, and
measure the query execution time for different input sizes.
The query contains both an aggregation and a self-join, and the runtime of these
expensive operators 
dominates all others.
Consequently, we expect the query to scale poorly when run entirely under MPC in
Sharemind, but for \sys to improve its performance by precomputing the
aggregation.
Figure~\ref{f:hhi} shows that Sharemind indeed takes over an hour to complete
the query at 100k input rows, while \sys scales roughly linearly in the
size of the input data.
This comes because \sys pushes the MPC frontier past aggregations for
the per-party revenue.
%
%
%
%
All data-intensive processing happens outside MPC in local Spark jobs, and only
a few records enter the final MPC, which consequently completes quickly.
Finally, we run the same query insecurely on the joint data using a single,
nine-node Spark cluster that processes all parties' combined data.
In this insecure setting, we observe similar performance.
Up to 10M, this insecure setup is slightly slower than \sys, as it runs one
job rather than three parallel jobs, but at 1.3B records, insecure Spark
benefits from the additional parallelism of the joint nine-node cluster and
completes quicker.
The market concentration query benefits from \sys pushing down the MPC
frontier and splitting the initial aggregation, but does not use hybrid
protocols.
%

\subsection{Hybrid operator performance}
\label{s:eval-ubench}

\begin{figure}[t]
 \centering
 \begin{subfigure}[b]{0.51\columnwidth}
  \setlength{\abovecaptionskip}{0pt}
  \setlength{\belowcaptionskip}{-2pt}
  \includegraphics{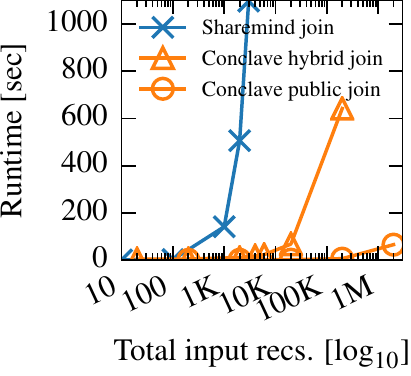}
  \caption{Hybrid join.}
 \end{subfigure}
 \begin{subfigure}[b]{0.47\columnwidth}
  \setlength{\abovecaptionskip}{0pt}
  \setlength{\belowcaptionskip}{-2pt}
  \includegraphics{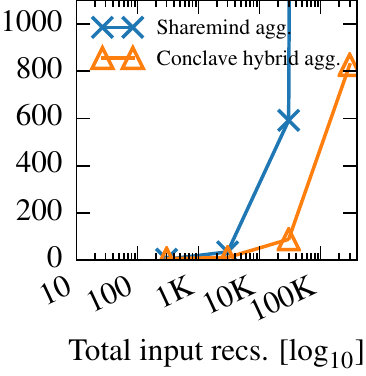}
  \caption{Hybrid aggregation.}
 \end{subfigure}
 \setlength{\belowcaptionskip}{-10pt}
 \caption{\sys's hybrid operators help scale joins and aggregations
   to large data by combining MPC and cleartext compute; At 10k records per party, Sharemind alone
   takes ten minutes (aggregation) and over twenty minutes (join).}
 \label{f:eval-ubench}
\end{figure}

%
%

%
\sys transforms queries to use hybrid protocols if a selectively-trusted
party participates (\S\ref{s:design-hybrid}).
We measure the impact of these hybrid protocols with queries with only
a single join or aggregation, and a growing amount of synthetic input
data.
We annotate the input relations' columns with STPs, allowing \sys to
apply hybrid operator transformations, and measure query runtimes.
We expect \sys's hybrid operators to reduce query runtimes, even though
the rewritten query contains additional operators.
Figure~\ref{f:eval-ubench} confirms this.
Without an STP, \sys must run the query entirely
under MPC, which exhibits performance similar to earlier benchmarks
(\S\ref{s:bg-scaling}).
\sys's hybrid join improves asymptotically over the MPC join; it replaces
$\mathcal{O}(n^{2})$ non-linear operations under MPC with oblivious
shuffles and indexing protocols, which require
$\mathcal{O}((n + m)\log{}(n + m))$ non-linear operations,
for input size $n$ and join output size $m$.
Consequently, using a hybrid join operator substantially improves
scalability: a hybrid join on 200k records takes just over ten minutes.
(At 2M input records, Sharemind runs out of memory while executing the
MPC part of the hybrid join.)

The public variant of the join operator scales even better since it
avoids the use of MPC altogether (hence, it completes at 2M records).
The public join's bottleneck is the local join.
The hybrid aggregation demonstrates similar speedups to the hybrid
join.
This is due to an asymptotic improvement: an oblivious shuffle with
$\mathcal{O}(n\log{}n)$ complexity replaces a sorting-network that
requires $\mathcal{O}(n\log^{2}n)$ comparisons.
In addition to the asymptotic improvement, the hybrid aggregation
also avoids oblivious comparison and equality operations, which are
slow in secret-sharing MPC.
%

\subsection{Credit card regulation query}
\label{s:eval-ssn}

\begin{figure}[t]
  \centering
  \includegraphics{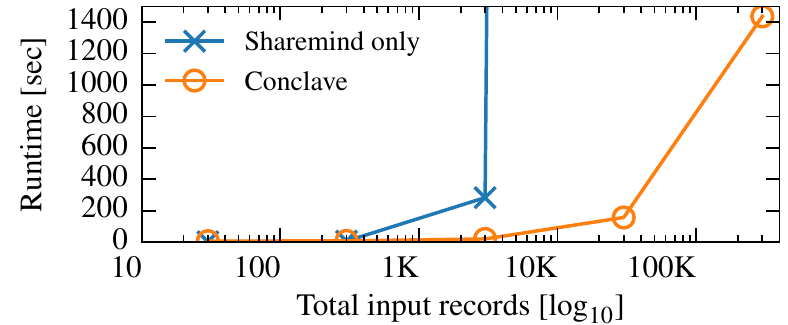}
  \setlength{\belowcaptionskip}{-2pt}
  \caption{\sys scales to two orders of magnitude more data on the
     credit card regulation query than pure Sharemind due to its
     hybrid join and aggregation.}
  \label{f:ssn}
\end{figure}

\sys's hybrid operators offer substantial benefits for queries whose
performance is dominated by aggregations and joins (a common case).
The credit card regulation query is an example: it first joins the
regulator's demographic information with the credit scores, and then
computes an aggregate (\viz the average score grouped by ZIP code).
The credit card companies trust the regulator, but not their competitors,
with the SSNs of their customers, and the regulator
wishes to keep the mapping from SSNs to ZIP codes private.
Hence, \sys can apply both the hybrid join and the hybrid aggregation
operator transformations to this query.
Even though these optimizations increase query complexity, we expect
them to reduce runtime, as both the join and the aggregation work can now
happen outside MPC.
Figure~\ref{f:ssn} confirms this.
Running the query entirely under MPC in Sharemind fails to scale beyond
3,000 total records; at 30k, the query does not complete
within two hours.
This comes despite the Sharemind baseline using a join implementation that
leaks output size, matching the leakage of \sys's hybrid join operator.
With \sys's hybrid operators, however, the query processes 300k records in under 25 minutes.
The experiment highlights that hybrid operators are crucial to obtaining
good performance for this query.
The query's first operator is a join, so \sys cannot push the MPC frontier
down and, without hybrid operators, would have to run the entire query
under MPC.
%

\subsection{Comparison with SMCQL}
\label{s:eval-smcql}

Finally, we compare \sys with the most similar state-of-the-art system,
SMCQL~\cite{smcql}.
SMCQL and \sys make different, complementary optimizations to speed up
MPC on for relational queries.
To compare, we configure \sys to be as similar to SMCQL as possible.
We disable the MPC frontier pushdown past local filters over private
columns (to match SMCQL's security guarantee), and manually implement
SMCQL's optimizations, which compose with \sys's optimizations.

SMCQL relies on the ObliVM~\cite{oblivm} garbled-circuit framework for
MPC.
ObliVM supports only two-party computations and is slower than
Sharemind, particularly on large data.
This difference is not fundamental: SMCQL could generate code for
Sharemind.
\sys uses the Sharemind backend in these experiments, and has two parties
provide input, while the third participates in the MPC without inputs.
\sys and SMCQL make comparable security guarantees, with the exception that
their backends' corruption thresholds differ (ObliVM: one of two,
Sharemind: one of three).
Because SMCQL requires more memory, the experiments use larger VMs with
32 GB RAM and 8 vCPUs.
%


\begin{figure}[t]
 \centering
 \begin{subfigure}[b]{0.51\columnwidth}
  \setlength{\abovecaptionskip}{0pt}
  \setlength{\belowcaptionskip}{-2pt}
  \includegraphics{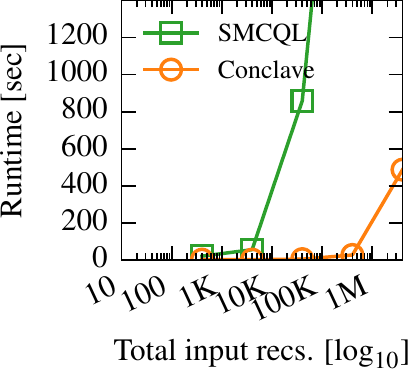}
  \caption{Aspirin Count.}
  \label{f:smcql:aspirin}
 \end{subfigure}
 \begin{subfigure}[b]{0.47\columnwidth}
  \setlength{\abovecaptionskip}{0pt}
  \setlength{\belowcaptionskip}{-2pt}
  \includegraphics{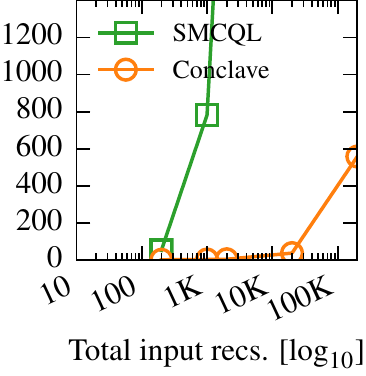}
  \caption{Comorbidity.}
  \label{f:smcql:comorbidity}
 \end{subfigure}
 \setlength{\belowcaptionskip}{-10pt}
 \caption{\sys outperforms SMCQL on the \emph{aspirin count}
    and \emph{comorbidity} queries~\cite[\S2.2.1]{smcql}.
    For \emph{aspirin count}, \sys's optimizations lift additional work out of MPC.
    At 200k records, SMCQL ran for over an hour for both queries.}
 \label{f:smcql}
\end{figure}

We benchmark the \emph{aspirin count} and \emph{comorbidity} queries from the SMCQL
paper.
We omit running the third query (\emph{recurrent c. diff.}), as \sys does not yet
support window aggregates, but summarize the performance we would expect.

\para{Aspirin Count~\cite[\S2.2.1]{smcql}.}
%
%
The query joins two input relations, \code{diagnoses} and \code{medications}
on public, anonymized patient IDs, filters by patient diagnosis and prescribed
medication (both private columns), and counts the results.
These inputs to the join are partitioned across two hospitals, \ie each party
holds part of \code{diagnoses} and part of \code{medications}.
SMCQL's ``slicing'' partitions the data on the public patient ID column.
Slices with patient IDs only one party has are processed locally at that
party, while the other slices must be processed under MPC.
For this experiment, we manually implement SMCQL's slicing and combine it
with \sys's public join.
%
%
We generated input data with a 2\% overlap between the parties'
uniformly-random patient IDs, similar to the HealthLNK
data~\cite[\S7.2]{smcql}.
We measure query runtime for increasing numbers of input records per party.
%

%
Figure~\ref{f:smcql:aspirin} shows that \sys consistently outperforms
SMCQL, and that it scales better.
At 40k rows, \sys completes in 3.7 seconds, while SMCQL takes 14.3 minutes;
SMCQL does not finish within an hour for 400k or more rows, while \sys takes
under two minutes.
Finally, \sys processes 4M input records in 8 minutes.\footnote{Bater
  \emph{et al.}\ also
  benchmarked \emph{aspirin count} for larger inputs of 42M \code{diagnoses}
  and 23M \code{medications} using eight servers to parallelize sliced
  MPCs, taking 23 hours to complete the query~\cite[\S7.3]{smcql}.
  \sys can likewise run additional Sharemind servers, but we lacked the
  resources to do so. We expect \sys to still outperform SMCQL on the full
  data set.}
This improvement is due \sys's public join and its sort elimination
optimization.
By combining the public join with slicing, \sys can compute the initial
join in the clear, and sends only rows for patient IDs present at both
parties into MPC.
By contrast, SMCQL still runs the join and the subsequent operations
obliviously for each private slice, which has quadratic cost in the size of
the slice.

Sort elimination allows \sys to avoid an expensive oblivious sort step otherwise
required for the distinct count.
\sys performs the sort in the clear, as part of the public join; since all operations under MPC are order-preserving no subsequent sorts are necessary.
This reduces the complexity of the MPC from
$\mathcal{O}(n\log{}n)$ to $\mathcal{O}(n)$.
SMCQL could likewise benefit from this optimization.

\extendedtr{
  \paragraph{Comorbidity.}
  The comorbidity query determines the most common conditions that affect patients suffering from \emph{c. diff.} (~\cite[\S2.2.1]{smcql}).
  It consists of a group-by aggregation on a private column (the patient's diagnosis) and an order-by (with a limit clause) of the result.
  As with aspirin count, the data is drawn from the HealthLNK dataset~\cite{healthlnk}.
  The input relation \code{diagnoses} is partitioned horizontally across two parties.

  \sys and SMCQL both split the aggregation into a local and an MPC step via an MPC-pushdown, and execute the rest of the query under MPC.
  The local aggregation reduces the amount of data entering MPC since neither \sys nor SMCQL pad the result.
  As such, the runtime of the query depends on the number of distinct keys in the patient diagnosis column.
  In our experiments, we set the number of distinct keys to 10\% of the total number of input rows.
  We increase the number of records in the input relation\,---\,\ie the
    total input is twice the $x$-axis value\,---\,and measure
  query runtime.

  As per Figure~\ref{f:smcql:comorbidity}, \sys outperforms SMCQL, with the gap increasing with data size.
  \sys scales to 200k total rows (20k rows entering MPC) while SMCQL takes over an hour at the 20k mark.
  As \sys and SMCQL apply the same optimizations, the improvement is due to a difference in MPC backends.
  \sys's Sharemind backend is better suited for arithmetic-heavy relational queries and considerably outperforms ObliVM~\cite{oblivm}.
  The relative performance improvement and scalability of \sys is lower than for aspirin count since the query does not lend itself to \sys's additional optimizations and requires two expensive operations (aggregation and order-by) to run under MPC.
}{
\para{Comorbidity.}
For this query, \sys and SMCQL apply identical optimizations (\viz MPC-pushdown),
and \sys's improves over SMCQL in Figure~\ref{f:smcql:comorbidity} only because
of its more efficient MPC backend (Sharemind vs.\ ObliVM).
%
%
%
%
}
\extendedtr{
\paragraph{Recurrent c.~diff.}
The third query in the SMCQL paper, \emph{recurrent c. diff.} consists of a
window aggregate, a join on public columns (patient IDs), and a selection of
distinct patient IDs from the result.
SMCQL performs the window aggregate and join in sliced mode, and runs the
distinct operator in the clear.
Like in \emph{aspirin count}, the query's performance is bottlenecked on the
join~\cite[\S7.3]{smcql}.
This query is amenable to \sys's public join optimization, so we expect \sys
to outperform SMCQL on the MPC steps due to \sys's use of Sharemind.
Based on our results for aspirin count (cf. Figure~\ref{f:smcql:aspirin}),
we expect \sys's public join (in combination with slicing) to match or
exceed the performance gain SMCQL realizes through sliced-mode execution.
}{

\para{Recurrent c.~diff.}
This query is amenable to \sys's public join optimization, but \sys
lacks a windowed aggregation operator it needs.
Based on our results for aspirin count (cf.\ Figure~\ref{f:smcql:aspirin}),
we expect \sys's public join (in combination with slicing) to match or
exceed the performance from SMCQL's sliced-mode execution.
}

%% file: related.tex
\section{Related Work}
\label{s:related}

We now highlight elements of \sys that relate to other approaches to
building privacy-protecting systems.
We omit prior work in MPC algorithms, frameworks, and deployments already
discussed in \S\ref{s:bg}.
Instead, we survey efforts that have made innovations in
``mixed mode'' operations, query rewriting, and query scalability for MPC.
%


%
\para{MPC with mixed mode operation.}
Wysteria~\cite{wysteria} performs mixed mode computations that move between MPC and local work.
Wysteria programs are written in a DSL that creates the programmer illusion of a
single thread of control.
However, the programmer must still manually annotate which blocks of the computation run under MPC, and which blocks are to be carried out locally.
These annotations are much more fine-grained than Conclave's input annotations, and require MPC proficiency.
By contrast, Conclave targets data analysts with minimal MPC knowledge, and focuses on automation. Conclave automatically optimizes queries to reduce the use of MPC, for instance via the MPC pushdown, whereas Wysteria requires the programmer to manually identify and implement such optimizations via the aforementioned per-block annotations.
Furthermore, \sys supports parallel cleartext processing using existing frameworks like Spark.
\para{Query rewriting and MPC alternatives.}
There are several efforts to optimize MPC via query rewriting, like \sys
does, at different levels of abstraction.
Kerschbaum \cite{kerschbaum-rewriting} operates at the circuit level,
transforming a manually assembled circuit into a different one with faster
execution under MPC (\eg using the distributive law to reduce the number of
multiplications).
Other systems perform query rewriting at the relational algebra level, such as
SMCQL~\cite{smcql} and Opaque~\cite{opaque}.
SMCQL, like \sys, uses column-level annotations, but differentiates only between
public and private columns.
However, SMCQL shares some optimization, such as parts of the MPC frontier
push-down, with \sys, and supports other complementary optimizations (slicing and
secure semi-joins).
\sys's annotations are more expressive, and \sys's hybrid protocols allow for
additional performance improvements.
Opaque, by contrast, runs most computation in the clear inside a protected
Intel SGX enclave as an alternative to MPC; its query rewriting focuses on reducing the number of
oblivious sorts required in distributed computation across multiple SGX
machines.
Similarly, Prochlo~\cite{bittau2017prochlo} combines the use of SGX, secret-sharing based techniques, and differential privacy~\cite{Dwork:2008:DPS:1791834.1791836} to implement large-scale application usage monitoring.
In contrast to \sys, Prochlo targets a setting where a large number of parties, \ie users, contribute data to the computation.
It furthermore relies on specialized hardware, and does not match MPC's full privacy guarantees.
%

%

%
\para{Protected databases and scalability.}
The protected database community has produced decades of research on scaling
secure query execution to the gigabyte-to-terabyte
range~\cite{BoschHJP14,HSSVYY16,FullerVYSHGSMC17}.
This includes work on optimizations for boolean keyword
search~\cite{PappasKVKMCGKB14,FaberJKNRS15}, as well as large, general subsets
of relational algebra \cite{KamaraM16}.
These works largely target querying a single protected database, as opposed
to \sys's distributed scenario.
%
%
%
%
Investigations into the scalability of secure MPC often involve laborious hand-optimization by groups of cryptographers on specific queries like set intersection \cite{ion2017private,
kamara2014scaling}, linear algebra~\cite{scalable-linear-reg-mpc}, or matching \cite{DoernerES16}.
%



%
\para{Inference and privacy.}
MPC protects sensitive state during computation but provides no restriction on
the ability to \emph{infer} sensitive inputs from the provided outputs.
Differential privacy (DP)~\cite{Dwork:2008:DPS:1791834.1791836} provably
ensures that the output of an analysis reveals nothing about any individual
input, but often uses a trusted curator to perform the analysis.
Several prior systems have combined MPC and DP to avoid the threat of parties
jointly reconstructing sensitive input data.
DJoin~\cite{djoin} does so for SQL-style relational operations (with query
rewriting, but without \sys's automation and hybrid protocols),
DStress~\cite{dstress} does so for graph analysis, and He \emph{et
al.}~\cite{HeMFS17} do so for private record linkage.
\sys does not currently leverage DP, but adding it would require no
fundamental changes to the query compilation.
%

%% file: concl.tex
\section{Conclusion and future work}
\label{s:concl}

\sys speeds up secure MPCs on ``big data'' by rewriting queries to
minimize expensive processing under MPC.
\sys runs queries in minutes that were impractical with previous MPC
frameworks or would have required domain-specific knowledge to
implement.
In the future, we plan to integrate other MPC backends into \sys, and
to make \sys choose the most performant MPC protocol for a query.
We are also interested in whether verifiable computation techniques can
be combined with \sys, and whether \sys can use adaptive padding to
avoid leaking relation sizes on the MPC boundary.
%


%
\sys is open-source and available at:
\begin{center}
  \textbf{\href{https://github.com/multiparty/conclave}{\tt https://github.com/multiparty/conclave}}.
\end{center}

%% file: acks.tex
\section*{Acknowledgements}

We thank Ran Canetti, Tore Kasper Frederiksen, Derek Leung, and Nickolai Zeldovich
for their helpful feedback on drafts of this paper.
We are also grateful to the helpful comments we received from our
anonymous reviewers, and our shepherd, Christian Cachin.
This work was funded through NSF awards CNS-1413920, CNS-1414119,
CNS-1718135, and OAC-1739000, and by the EU's Horizon 2020 research and innovation programme under grant agreement 731583.
%

%% file: appendix.tex
\appendix

\section{Security analysis}
\label{s:appendix}

To prove \sys's security guarantees, we analyze the security of \sys's push-down, push-up, and hybrid transformations individually and collectively.

\mv{(2/1, for me to do later) Consider applying differential privacy to the cardinalities of intermediate relations. They are essentially ``count'' queries, for which DP is well-understood.}

\subsection{Definitions}
A secure MPC protocol $\pi$ \emph{securely computes} function $f$ subject to a (monotone-decreasing) adversary structure \Structure if $\pi$ enables parties with inputs \Inputs to learn $f(\Inputs)$ while ensuring that no sufficiently small colluding subset of parties can learn any additional ``useful'' information from everything they view during the protocol.
We codify this statement in pieces.

First, an adversarial set of colluding parties \Corrupt is deemed to be \emph{permissible} if $\Corrupt \in \Structure$; these are the only adversaries that $\pi$ vouches to withstand.
Second, the \emph{view} of an adversary is defined as the state of all colluding parties in \Corrupt together with the set of all network messages they observe.
Third, we simultaneously guarantee MPC's correctness and security by requiring that \Corrupt's view can be simulated given only the colluding parties' inputs ($\CorrInputs = \{x_i : i \in \Corrupt\}$) and the size of all parties' inputs (\UncorrLengths, where $\ell_i = \abs{x_i}$); ergo, \Corrupt cannot learn more information than this from its own view.
Formally, we require that for all permissible adversaries \Corrupt, there exists a simulator \Sim such that
the following two ensembles are computationally indistinguishable \cite{Goldreichv1} when the distinguisher and \Sim run in time polynomial in the security parameter:
\begin{equation}
\langle \CorrInputs, \UncorrLengths, \out, \view \rangle
\approx
\langle \CorrInputs, \UncorrLengths, f(\vec{x}), \Sim(\CorrInputs, \UncorrLengths, [f(\vec{x})])
  \rangle
\text{.}
\label{eq:MPC}
\end{equation}
Correctness follows because $\out = f(\Inputs)$ and security stems from the fact that the adversary's view contains ``no more information'' than the adversary's own input and the length of the honest parties' inputs.
If the adversarial set \Corrupt includes the receiving party, then \Sim is also given $f(\Inputs)$ since the adversary is supposed to learn it through the execution of $\pi$.

\nv{Mention in the body of the paper that \sys doesn't currently support reactive queries.}

\subsection{Semi-Honest Security without Hybrid Operators}

Without hybrid operators, the execution of \sys on some function $g$ proceeds in three phases as shown in Figure~\ref{f:analysis-rewrite-pipeline}.
\begin{itemize}
    \item Local pre-processing $d$: each party calculates $d_i(x_i)$.
    \item A single MPC calculation over a set of operators $f$ that ultimately reveals some value $y$ to the receiving party.
    \item Local post-processing $u$: receiving party outputs $z = u(y)$, where $u$ is an invertible function.
\end{itemize}
\sys's push-down and push-up transformations choose the local pre- and post-processing operators.

\extendedtr{
The security of this transformation can be proved via a standard composition argument.
We use the composition lemma from Goldreich~\cite[\S 7.3.1]{DBLP:books/cu/Goldreich2004}.

\begin{lemma}%
\label{lem:composition}
Suppose there exists an MPC protocol \piF that securely computes $f$ and another protocol \piGF that securely computes $g$ but is permitted to make oracle queries to $f$, where both protocols provide semi-honest security with respect to the same adversary structure \Structure.
Then, $g$ can be securely computed (without oracle queries) via the protocol \piG that runs \piGF but substitutes every oracle call to $f$ with an execution of \piF; this protocol leaks the length of the input to $f$ in the process.
\end{lemma}

This composition statement applies to all secure computation protocols, whether based on secret sharing or garbled circuits. Ergo, the lemma applies to all \sys backends.
} 
{The security of this transformation can be proved via a standard composition argument that applies to all secret sharing or garbled circuit-based MPC protocols.
}

\begin{theorem}
\label{thm:composition-without-hybrids}
Consider \sys with an MPC backend that provides semi-honest secure computation under adversary structure \Structure.
Suppose \sys splits the function to be computed $g = u \circ f \circ d$ into a set of local per-party pre-processing operators $\{d_i\}$, an operator $f$ intended for secure computation, and a local post-processing operation $u$ performed by the receiving party.
Then, \sys securely computes $g$ as per equation \eqref{eq:MPC} with the same adversary structure \Structure, except that the protocol reveals the input lengths of $f$ and not $g$ (i.e., $\ell_i = \abs{d_i(x_i)}$).
\end{theorem}

\extendedtr{
For simplicity, we assert without proof that \sys's query transformations preserve correctness (mostly due to the distributive law). The proof below focuses on security.

\begin{proof}
In order to apply Lemma \ref{lem:composition} to $g = u \circ f \circ d$, we must show the existence of two MPC protocols \piF and \piGF.

\sys's existing backend already provides us with a protocol \piF that securely computes $f$ under some given adversary structure \Structure.
Furthermore, we can directly construct a protocol \piGF that securely computes $g$ under the same adversary structure \Structure given oracle access to $f$.
This protocol \piGF proceeds as follows.
First, each party locally computes its own $d_i(x_i)$ operation and feeds the result to the $f$ oracle.
Second, the receiving party takes the oracle's response $y$ and calculates the output $z = u(y)$.

We show how to simulate the view of any adversarial coalition $\Corrupt \in \Structure$.
If \Corrupt does not include the receiving party, then the adversary never receives any messages and the only intermediate state it computes is $d_\Corrupt = \{d_i(x_i) : i \in \Corrupt\}$, which the simulator can also compute directly from its inputs.
If \Corrupt includes the receiving party, then the only additional item in the view is $y$, which the simulator can reconstruct as $u^{-1}(g(\Inputs))$ since $u$ is invertible and the simulator for this coalition is given $g(\Inputs)$.

Finally, with protocols \piF and \piGF constructed, we may apply Lemma \ref{lem:composition} to produce a protocol \piG that securely computes $g$.
Note that equation \eqref{eq:MPC} permits the simulator for \piF to use the input lengths to $f$, whereas the simulator for \piGF never utilized any input length information. Ergo, simulation of \piG only requires the input lengths to $f$.
\end{proof}
}{
\begin{proof}
See \fullversion~\cite{conclave-tr}.
\end{proof}
}

\subsection{Hybrid Operators Stand-Alone Security}
\label{ss:hybrid-ops-security}

\mv{Modify theorem statements in this section to say that Conclave's permissible sets do not allow collusion between the STP and the regular parties (even if the original permissible set \Corrupt does).}

In this section, we provide simulation-based proofs of semi-honest static security for
\extendedtr{
the three hybrid operators described in \S \ref{s:design-hybrid}.
Each proof is stand-alone; that is, it presumes that the entire \sys calculation contains a single operator.
Once again, we assert correctness without proof, and we focus solely on the simulation security arguments.
}{
the hybrid join and hybrid aggregation protocols (\S\ref{s:design-hybrid}).
}
The proofs are partitioned based upon whether the colluding set includes the selectively-trusted party (STP). If so, then recall that the STP must operate alone; it is not permitted to collude with other participants.
\extendedtr{}{Our extended technical report~\cite{conclave-tr} includes full proof details and discussion of the public join.}

\begin{theorem}
\label{thm:hybrid-standalone}
Suppose that \sys uses a secure MPC backend
whose adversary structure \Structure includes $\{\text{STP}\} \in \Structure$ but $\text{STP} \notin \delta$ for any other set $\delta \in \Structure$, and it supports secure protocols for oblivious shuffles and indexing.
Then the algorithm for each hybrid operator in \S \ref{s:design-hybrid} results in a standalone secure computation of its corresponding operator as per equation \eqref{eq:MPC}, subject to the following leakage: the STP learns the join or aggregation key column, and all parties learn the number of rows for the input and result of the operator.
\end{theorem}

%

\nv{(2/1) Changed} \mv{(2/1) Even more changes}

\extendedtr{
Because the theorem focuses on standalone security, we demonstrate it separately for each of the 3 hybrid operators.

\begin{proof}[Proof for hybrid join]
First, we analyze security against the STP.
In step \ref{enum:hybrid-join-stp} of the algorithm, the STP receives the set of key columns corresponding to the join operator, which the simulator can emulate because it receives this information as leakage.
In all other steps, the STP is either not involved or it participates as a single party within an MPC protocol, which is simulatable because the STP on its own is a permissible set.

Second, we construct a simulator \Sim for a permissible coalition of regular parties $\Corrupt \in \Structure$.
This coalition participates in a secure computation involving the other parties' input data along with (in steps \ref{enum:Hjoin-regular-first}--\ref{enum:Hjoin-regular-last}) the STP's mapping of index relations between parties.
Because \Corrupt is a permissible set, the \sys MPC backend guarantees the existence of a simulator \Sim' that can emulate \Corrupt's view when provided with \Corrupt's inputs \CorrInputs as well as the input lengths for both the honest parties' relations \UncorrLengths and the STP's index-mappings $\ell_{\text{STP}}$.
Since \Sim receives \CorrInputs and \UncorrLengths by definition and can compute $\ell_{\text{STP}}$ from its leakage (the eventual number of rows in the joined result), \Sim can execute \Sim' to emulate \Corrupt's view.
\end{proof}

\begin{proof}[Proof for public join]
Recall that the public join operator is not conducted under MPC at all: the parties disseminate their key columns in the clear and the aiding server calculates the join in the clear as well.
This operator requires a very large amount of leakage: every participant agrees to disclose the relevant key columns to all other participants.
Given such leakage, the view of a public join can be trivially simulated by re-calculating the operator in the clear following the same procedure as the aiding server.
\end{proof}

\begin{proof}[Proof for hybrid aggregation]
This proof is similar to that for the hybrid join.
For the STP: the view of the message it receives in step \ref{enum:hybrid-agg-stp} can be simulated by taking a random permutation of the multiset of group-by key values that it receives as leakage.
In the remaining steps, the STP merely participates in the MPC protocol so its view is simulatable because the STP on its own is a permissible coalition.

Generating a simulator \Sim for a permissible coalition of regular parties $\Corrupt \in \Structure$ once again relies upon executing the simulator \Sim' provided by the MPC backend for the MPC steps of the aggregation (steps \ref{enum:Hagg-regular-first}--\ref{enum:Hagg-regular-last}), and again the only challenge is to ensure that \Sim can calculate the length information to provide to \Sim'.
In a hybrid aggregation, the length of the STP's input in this protocol only depends on the input lengths by the regular parties, and the length of the STP's output equals the output length of the entire aggregation.
As a result, \Sim and \Sim' require precisely the same length information.
\end{proof}
} 
{ 
\begin{proof}[Proof sketch]
The STP only receives information in one step of each algorithm (step \ref{enum:hybrid-join-stp} of the hybrid join or step \ref{enum:hybrid-agg-stp} of the hybrid aggregation), and the simulator can emulate this message by randomly permuting the join/aggregation key columns it receives as leakage.
Additionally, we can simulate the view of a colluding set of regular parties in the MPC steps of the protocols (steps \ref{enum:Hjoin-regular-first}--\ref{enum:Hjoin-regular-last} for hybrid join or steps \ref{enum:Hagg-regular-first}--\ref{enum:Hagg-regular-last} for hybrid aggregation) by running the simulator provided by \sys's MPC backend. We must provide the backend simulator with the lengths of intermediate state during the MPC steps of the hybrid operator, which we can calculate from the length leakage stated within the theorem.
\end{proof}
} 

\subsection{Semi-Honest Security with Hybrid Operators}

To complete the security analysis of \sys in the semi-honest setting, we augment the composition argument to account for the presence of hybrid operators.
In its full generality, \sys's static analysis splits the function $g$ to be computed into the following series of $2k+1$ functions:
\begin{equation}
\label{eq:composition-hybrids}
g = u \circ e^{k} \circ e^{k-1} \circ \cdots \circ e^2 \circ f^2 \circ e^1 \circ f^1 \circ d \text{,}
\end{equation}
where $d$ and $u$ denote local pre- and post-processing, the $f^j$ functions are intended for (generic) secure computation using the \sys backend, and the $e^j$ functions are hybrid operators.
We assume without loss of generality that the distributed computation begins and ends with non-hybrid MPC steps, which might be identity function operators.

To invoke composition when several operators in a row are processed using secure computation, we ``lift'' the functions to their equivalents that operate over shared state. That is:
\begin{itemize}
    \item $\hat{d}$ = local pre-processing and randomized sharing.
    \item Each $\hat{e}^{j}$ and $\hat{f}^{j}$ receives a sharing of the corresponding input for $e^j$ or $f^j$, respectively, and produces a random sharing of the corresponding output.
    \item $\hat{u}$ = reconstruction and local post-processing by the receiving party.
\end{itemize}
%
%
\extendedtr{
When we say ``sharing'' here, we use the term generically in order to consider both secret sharing-based MPC and garbled circuit-based MPC.
For the latter, we follow the approach of Demmler et al.~\cite{Demmler0Z15} to consider one bit of secret state in a garbled circuit as being ``shared'' by having the garbler know the values corresponding to the two wire labels and the evaulator know exactly one of the wire labels.
}{} 

Like most MPC protocols, the backends for \sys are \emph{reactive} and permit calculations directly over the shares.
That is, they can securely compute the $\hat{f}^j$ or $\hat{e}^j$ functions over an already-shared state, without the need to perform an initial sharing of inputs or reconstruction of the output.
We use this observation in the following theorem.

\begin{theorem}
\label{thm:composition-with-hybrids}
Consider \sys with an MPC backend that provides semi-honest secure computation under adversary structure \Structure such that $\{\text{STP}\} \in \Structure$ and furthermore this is the only set in \Structure that includes the STP.
Suppose also that \sys's backend is reactive, with well-defined algorithms to share and reconstruct state and with all intermediate state during a calculation being shared among the participants.
Given a function $g$ that \sys partitions into the sequence of operators in equation \eqref{eq:composition-hybrids}, \sys securely computes $g$ as per equation \eqref{eq:MPC} in the server-aided setting with adversary structure \Structure subject to the following leakage: every party learns the input and output lengths of $d$ and all $e^j$, and the regular and selectively-trusted parties receive the stand-alone leakage of each constituent hybrid operator $\hat{e}^j$.
\end{theorem}


\extendedtr{
As before, we presume \sys preserves correctness and focus on proving security of the composed construction.
\begin{proof}
Theorem \ref{thm:hybrid-standalone} provides us with MPC protocols for each $\hat{f}^j$ and $\hat{e}^j$. Additionally, we can construct secure computation for $\hat{d}$ and $\hat{u}$ just as was done in the proof of Theorem \ref{thm:composition-without-hybrids} with minor changes:
simulating the initial sharing step within $\hat{d}$ requires generating shares of $d_i(x_i)$ for $i \in \Corrupt$ and shares of random values for the parties outside of the coalition,
and inverting $\hat{u}$ requires computing $u^{-1}$ and then generating random shares of the result.

All that remains is to apply Lemma \ref{lem:composition}.
We observe that the simulator for each $\hat{f}^j$ or $\hat{e}^j$ operator is provided with the length of its own input, plus optionally the leakage stated for each hybrid operator $\hat{e}^j$.
We have already provided the overall simulator \Sim for \piG with all of the length information as leakage; note in particular that the output length of $\hat{f}^j$ equals the input length of $\hat{e}^j$ and the input length of $\hat{f}^j$ equals the output length of $\hat{e}^{j-1}$ (or $\hat{d}$ in case $j=1$).

Ergo, \Sim can emulate the view during an execution of $g$ by executing all of the simulators for $\pi^{\hat{d}}$, $\pi^{\hat{f}^1}$, $\pi^{\hat{e}^1}$, $\ldots$, $\pi^{\hat{u}}$ in sequence.
We can show that \Sim is indistinguishable from the real world via a simple hybrid argument, where in the $i^{\text{th}}$ hybrid for $i \in \{0, 1, \ldots, 2k+1\}$, the view of the first $i$ operators is real and the view of the remaining operators is simulated. Each pair of adjacent hybrids is indistinguishable due to the simulation security of the $i^{\text{th}}$ operator, and the endpoints correspond to the real and simulated views.
\end{proof}
}{
\begin{proof}
See \fullversion~\cite{conclave-tr}.
\end{proof}
}

The union of all leakages stated in Theorem \ref{thm:composition-with-hybrids} can potentially be large; its potential for harm in practice depends strongly on the parties' privacy concerns about their input data, as well as on the number and type of hybrid operators invoked.
That having been said, our annotation propagation method ensures that even the composition of all leakages of intermediate state to the STP is no worse than simply revealing all of the columns that the regular parties are willing to share with the STP. %
That is, applying our recursive definition of trust sets to Theorem \ref{thm:composition-with-hybrids} yields the following corollary.

\begin{corollary}
\label{cor:annotation-propagation}
\nv{Discuss what should go in the camera-ready for this section.}
Suppose \sys and its MPC backend satisfy all of the requirements listed in Theorem \ref{thm:composition-with-hybrids} and that \sys is tasked to compute function $g$ as given in equation \eqref{eq:MPC}.
Then, \sys securely computes function $g$ subject to the following leakage: every party learns the input and output lengths of $d$ and all $e^j$, and the STP receives the contents of all input columns annotated by the regular parties.
\end{corollary}

%


\extendedtr{
\begin{proof}
The only difference between this theorem and Theorem \ref{thm:composition-with-hybrids} is the leakage to the STP. We assert without proof the correctness of \sys's manual specification of the column dependency for each operator, and focus on security.

\sys deploys a hybrid join (respectively, hybrid aggregation) only if the STP is in the trust set of the key (resp., group-by) column of the hybrid operator's result.
%
Based on \sys's recursive definition of the ``trust set'' of an intermediate operator, it follows that there exists a propagation algorithm \Propagate that, when given all input data that has been annotated for sharing with the STP, can calculate sufficiently many columns of the DAG to derive all join key (respectively, group-by) columns of the result and all operands for the hybrid join (resp., hybrid aggregation).
Hence, we can construct a simulator \Sim to satisfy this theorem simply by running \Propagate on the leaked input data to reconstruct the leakage required to run the simulator within Theorem \ref{thm:composition-with-hybrids}.
\end{proof}
}{
A proof of this claim, like many others in this appendix, can be found in \fullversion~\cite{conclave-tr}.
}

\extendedtr{
\subsection{Malicious Security without Hybrid Operators}
\label{ss:malicious}

Unlike the rest of this paper, in this section we briefly and informally examine \sys's ability to withstand malicious attack. With a few simple modifications, \sys can provide malicious security up to abort and without guaranteed output delivery or fairness.

To begin, we examine where a malicious actor \Adv could render moot the analysis of Theorem \ref{thm:composition-without-hybrids} by deviating from the secure computation of $g = u \circ f \circ d$.
There exists a more complex version of Lemma \ref{lem:composition} that provides malicious security as long as both \piF and \piGF can withstand malicious attacks \cite[\S 7.4.2]{DBLP:books/cu/Goldreich2004}.
To compute $f$ securely, \sys simply needs to include a backend that withstands malicious actors.
Ergo, the only remaining question is the security of \piGF (i.e., the calculations of $u$ and $d$) under malicious attack.

For the post-processing operator $u$: if \Adv doesn't corrupt the receiving party then she has no way to influence $u$. If \Adv does corrupt the receiving party then she can influence $u$ arbitrarily or refuse to calculate it altogether, thereby eliminating any chance of guaranteeing output delivery or fairness. However, because $u$ is a local computation any of these deviations must be simulatable so \Adv cannot learn any new information from deviating at this stage.

For the pre-processing operator $d$: \Adv could potentially learn information if a malformed calculation of $d_i(x_i)$ is fed into subsequent operators, so we must ensure that the local operators are calculated correctly.
Because each party has the right to choose its own input $x_i$, it suffices to ensure that its calculation of $x_i' = d_i(x_i)$ is \one{} independent from the other parties' pre-processing calculations and \two{} a value in the support of $d_i$. These conditions can be satisfied with a commitment scheme and a zero knowledge proof, respectively.

In summary, \sys can provide malicious security if:
\begin{enumerate}
    \item It uses a malicious secure backend,
    \item It adds an initial round of communication during which each party commits to its local pre-processing output.
    \item It ties $d$ and $f$ together by having each party prove in zero knowledge that its contribution to \piF equals the value previously committed and is in the range of $d_i$.
\end{enumerate}
We omit formal definitions and analysis.
The current \sys prototype does not yet support malicious security.
}{}